\documentclass[reprint,amsmath,amssymb,aps,superscriptaddress,tightenlines,article,twocolumn]{revtex4-2}
\usepackage{graphicx}
\usepackage{dcolumn}
\usepackage{bm}
\usepackage{epsfig,graphics,amsthm,amsmath,amssymb}
\usepackage{textcomp}
\usepackage{subfigure}
\usepackage{threeparttable}
\usepackage{csquotes}
\usepackage{color}
\usepackage{textcomp}
\usepackage{setspace}
\usepackage{dcolumn}
\usepackage[version=3]{mhchem}
\usepackage{lmodern}
\usepackage{microtype}
\usepackage{epstopdf}
\usepackage{upgreek}
\usepackage{siunitx}
\usepackage[colorlinks,citecolor=blue,linkcolor=red]{hyperref}

\theoremstyle{dgthm}

\theoremstyle{dgdef}

\newcommand{\abs}[1]{\left\vert#1\right\vert}

\renewcommand{\i}{\mathrm{i}}

\newcommand{\set}[1]{\left\{#1\right\}}

\newcommand{\bmE}{\boldsymbol E}

\newcommand{\bmG}{\boldsymbol G}

\newcommand{\bmP}{\boldsymbol P}

\begin{document}

\title{Revealing topological phase in Pancharatnam-Berry metasurfaces using mesoscopic electrodynamics}

\author{Zhanjie Gao}
\email{Zhanjie Gao, and Sandeep Golla contributed equally to this work.} 
\affiliation{State Key Laboratory of Precision Spectroscopy, East China Normal University, Shanghai 200062, China}
\author{Sandeep Golla}
\email{Zhanjie Gao, and Sandeep Golla contributed equally to this work.}
\affiliation{Universit\'{e} C\^{o}te d$^\prime$Azur, CNRS, CRHEA, rue Bernard Gregory, Valbonne 06560, France}
\author{Rajath Sawant}
\affiliation{Universit\'{e} C\^{o}te d$^\prime$Azur, CNRS, CRHEA, rue Bernard Gregory, Valbonne 06560, France}
\author{Vladimir Osipov}
\affiliation{State Key Laboratory of Precision Spectroscopy, East China Normal University, Shanghai 200062, China}\affiliation{Holon Institute of Technology, 52 Golomb Street, POB 305 Holon 5810201, Israel}
\author{Gauthier Briere}
\affiliation{Universit\'{e} C\^{o}te d$^\prime$Azur, CNRS, CRHEA, rue Bernard Gregory, Valbonne 06560, France}
\author{Stephane Vezian}
\affiliation{Universit\'{e} C\^{o}te d$^\prime$Azur, CNRS, CRHEA, rue Bernard Gregory, Valbonne 06560, France}
\author{Benjamin Damilano}
\affiliation{Universit\'{e} C\^{o}te d$^\prime$Azur, CNRS, CRHEA, rue Bernard Gregory, Valbonne 06560, France}
\author{Patrice Genevet}
\email{patrice.genevet@crhea.cnrs.fr}
\affiliation{Universit\'{e} C\^{o}te d$^\prime$Azur, CNRS, CRHEA, rue Bernard Gregory, Valbonne 06560, France}
\author{Konstantin E. Dorfman}
\email{dorfmank@lps.ecnu.edu.cn}
\affiliation{State Key Laboratory of Precision Spectroscopy, East China Normal University, Shanghai 200062, China}

\begin{abstract}
{Relying on the local orientation of nanostructures, Pancharatnam-Berry metasurfaces are currently enabling a new generation of polarization-sensitive optical devices. A systematical mesoscopic description of topological metasurfaces is developed, providing a deeper understanding of the physical mechanisms leading to the polarization-dependent breaking of translational symmetry in contrast with propagation phase effects. These theoretical results, along with interferometric experiments, contribute to the development of a solid theoretical framework for arbitrary polarization-dependent metasurfaces.}
\end{abstract}


\maketitle

\section{Introduction}
Pancharatnam-Berry (PB) metasurfaces, made of periodic arrangements of subwavelength scatterers or antennas, have been extensively studied over the last few years and are currently considered as a forthcoming substitute of bulky refractive optical components \cite{1,2}. The reflection and refractive properties of light at interfaces can be efficiently controlled by appropriately designing the phase profile of these surfaces \cite{3}. Several applications of PB metasurfaces, ranging from coloring to the realization of multifunctional tunable/active wavefront shaping devices, have been proposed\cite{4}. As a result of the fascinating degree of the wavefront manipulation offered by metasurfaces, this technology is currently bursting through the doors of industry, particularly driven by their potential application in redefining optical designs, such as lenses \cite{6,7,8,9}, holography \cite{10,11,12b}, polarimetry \cite{13,14,15b} and a variety of broadband optical components, including free-form metaoptics \cite{17,18,19,20,20b}.

Despite these applications, significant efforts are currently being made in deriving proper theoretical frameworks to guide the design of complex components.
Most of the disruptive attempts in controlling light-matter interactions rely on a fully vectorial Maxwell's equations, such as effective medium theories \cite{20c,20d,20e}, and the comprehensive understanding of their polarization responses generally obtained using extensive numerical simulations, such as finite element method \cite{Khanikaev} or finite-difference time domain techniques \cite{3,sunet,Vahabzadeh}, which often shows the quantitative simulation results but lacking of qualitative physical interpretations \cite{de2007colloquium,cz2013enhancement,Liu2017Momentum}. Another approach, Green's function method and diffraction theory for gratings, provides partial interpretation of a few diffractive properties of metasurfaces. The generalized Snell's law can be then understood as a maximum grating efficiency in a given diffraction order \cite{smith2011analysis,larouche2012reconciliation}. However, a vectorial theoretical framework is still required to clearly explain why the generalized Snell's law occurs in the cross-polarized transmitted fields in PB metasurface system in the -1st or 1st diffraction orders only.
To overcome these difficulties, the concept of geometric phase (PB phase), which is responsible for the conversion of the polarization state in the linearly birefringent medium  \cite{Pancharatnam1956,berry1987the,
kuratsuji1998maxwellsch,bliokh2006conservation,zhu2019generalized},
is introduced.
Several works have shown that the transmission matrix which describes the birefringent response can be separated into co-polarized and cross-polarized beams in the circular basis by applying the PB phase induced by the orientation of nano-antennas \cite{36,pauli_chinese,Huang}. However, this approach does not originate from first principle derivation and is not capable of explaining other diffractive properties of PB metasurfaces, such as the connection between generalized Snell’s law and polarization conversion. Obviously, each of these approaches just capture a part of the whole physical mechanism.
To fill the gap between these concepts and incomplete demonstrations, a theoretical framework is highly needed to interpret all the diffractive properties of PB metasurfaces in a precise and systematic way.

In this letter, we propose a systematic mesoscopic electrodynamical theory to study the polarization-dependent metasurface, showing that the transmission of a co-polarized beam only acquires global phase associated with the antenna response, called ``the propagation phase delay'',  while the transmission of a cross-polarized beam is sensitive to both PB and propagation phases. We extend this phase effect to a more general situation by decomposing the arbitrary polarization of a normally incident light in circular basis, showing that each eigenstate acquires an opposite phase delay due to the topological phase retardation associated with the PB phase (see Eq. \ref{eq:berryp}). Furthermore, we derive a fully electrodynamical expression and conduct optical measurements to analyze and validate this theoretical framework describing the diffractive properties of topological phase gradient metasurfaces \cite{Hasman_2002, Brongersma_2014}, including the physical mechanisms of the coexistence of the  zero and nonzero phase gradient leading to the ordinary and generalized Snell's law, and the universal principles of co-polarization and cross-polarization transmission.
\begin{figure}[!htbp]
	\centering
	\includegraphics[width=0.94\linewidth]{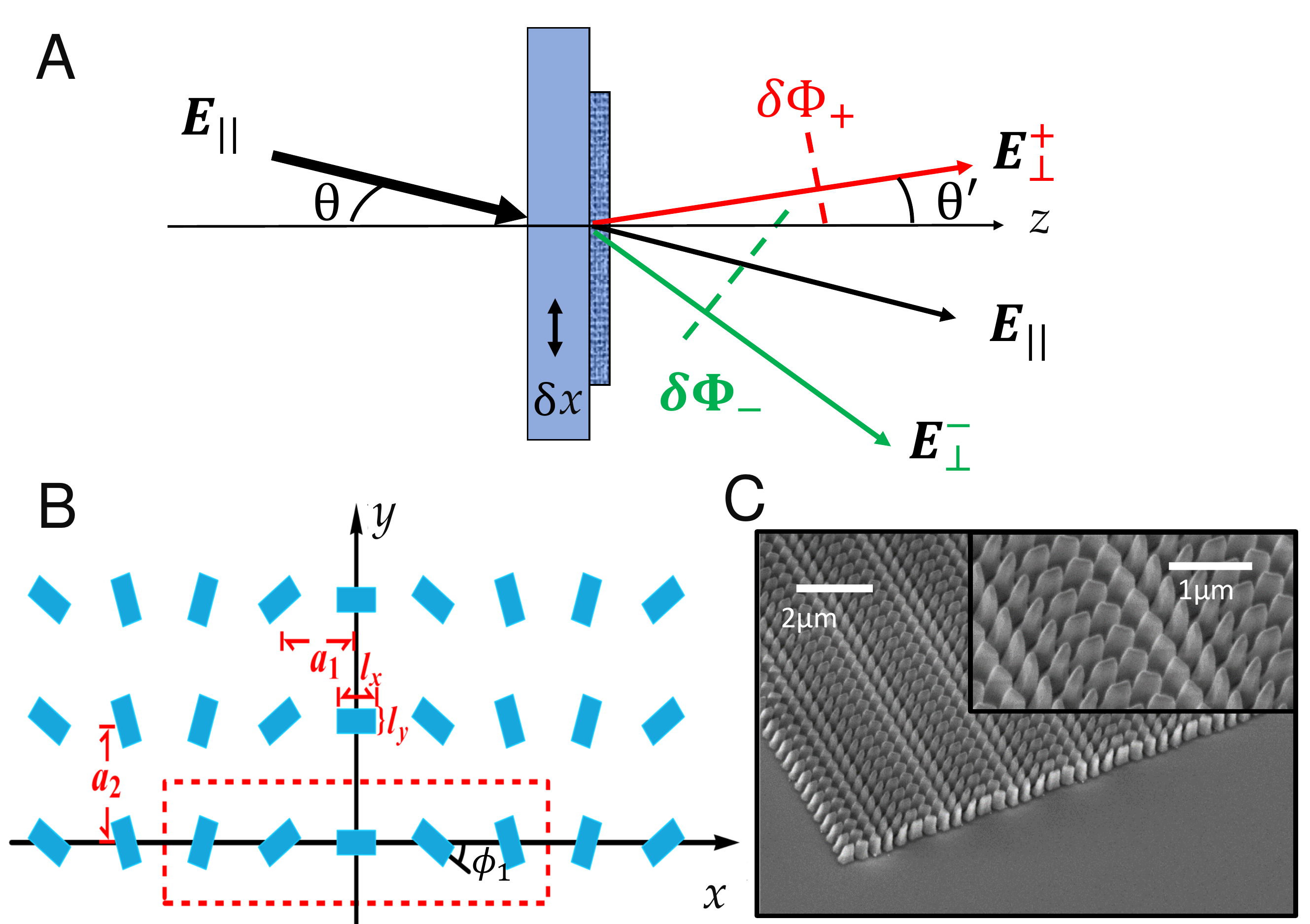}
	\caption{\footnotesize (A) Schematic explanation the transmission properties of the PB metasurface where $\mathbf{E}_{||}$ and $\mathbf{E}_{\bot}^{\pm}$ denote co-polarized and cross-polarized beams, $\theta$ and $\theta'$ are the incident and refraction angles, respectively. (B) Schematic of the nanostructure array used to generate both classical and anomalous refraction. The array consists of a repetition of a unit cell containing five rotated nanopillars with dimensions $l_x\times l_y\times l_z$ separated by subwavelength distances $a_1$ and $a_2$ along $x$ and $y$, respectively. The rotation angle $\phi_1=\frac{-\pi}{2N+1}$. (C) Scanning electron micrograph of a representative GaN-based PB metasurface. }
	\label{fig1}
\end{figure}

\section{ The Mesoscopic Model}

The topological phase occurring on the converted state of polarization is generated after transmission across a PB metasurface, as shown in Fig. \ref{fig1}A. To study these interface phenomena, we consider non-magnetic PB metasurfaces and express the transmitted light starting from Maxwell's equations for monochromatic light in the media~\cite{Landau} (CGS units)
\begin{eqnarray}\label{1.21}
\Delta\bmE-\frac{\partial^2}{c^2\partial t^2} \bm E = -4\pi\nabla(\nabla\cdot\bmP ) + \frac{ 4\pi\partial^2}{c^2\partial t^2} \bmP,\label{1.22}
\end{eqnarray}
where we assume that the electric field $\bmE$ with frequency $\omega_i$ is far detuned from any electric resonance and $\bmP$ denotes the polarization of the metasurface and substrates on both side (see Section S1.A in Supplementary Materials (SM) for more details).

The metasurface can be represented by a lattice with a primitive cell consisting of $2N+1$ Gallium Nitride (GaN) nanopillars distributed along the $x$ coordinate, which  corresponds to a reciprocal lattice vector $\bmG_{mn}=\frac{2\pi m}{(2N+1)a_1}\bm e_x+\frac{2\pi n}{a_2}\bm e_y$, where $m, n$ are integers, $a_1$ and $a_2$ are the nearest spacing between the individual nanopillars along $x$ and $y$ directions, respectively, see Fig. \ref{fig1}B. The linear response of the individual nanopillars, at position $j$ in the unit cell, is described by a polarization vector (see Section S1.B in SM for more details)
\begin{equation}
\begin{split}\label{eq:P0}
\bmP_j(z,\bm \rho,\omega)=N_0\sum_{m,n}\int_\mathbf{Q} d^2 \bm{\kappa}f_{mn,j}(\phi_j)\bmE(z,\bm\kappa,\omega)e^{\i \psi_{mn,j}}.
\end{split}
\end{equation}
Here $\psi_{mn,j}= \bmG_{mn}\cdot(\bm\rho-j\bm a_1)+\bm{\kappa}\cdot\bm \rho$ describes only the propagation phase, and the form-factor of the $j$-th element in the $mn$-th lattice unite cell is  $f_{mn,j}(\phi_j)=\tilde\Omega(\bmG_{mn})/[\pi(2N+1)a_1a_2]$ where $\tilde\Omega(\bmG_{mn})$ is the Fourier transform of geometric shape factor $
\Omega(\bm \rho)=\mathcal{H}(\abs {x}\le l_x/2)\mathcal{H}(\abs{y}\le l_y/2)$ with Heaviside function $\mathcal{H}(\text{condition})=\set{1,\mbox{ when condition is true;~} 0, \mbox{ when condition is false}}$. The momentum integration over $\bm\kappa$ runs over $\mathbf{Q}$ - the first Brillouin zone. The coefficient $N_0=\frac{\chi_0 }{8\pi^2}$ includes the nanopillars material susceptibility  $\chi_0$.

The translational symmetry of the metasurface dictates the form of the solution which is given by
\begin{equation}\label{eq:Esol}
\bmE(z,\bm\rho)=
\sum_{m,n}e^{\i\bmG_{mn}\cdot\bm\rho}\int_\mathbf{Q} \frac{d^2\bm \kappa}{(2\pi)^2} \bmE_{mn}(z,\bm \kappa) e^{\i\bm \kappa \cdot\bm \rho}.
\end{equation}
Considering the thickness $l_z$ to be much smaller than the $xy$ dimension of the metasurface, we neglect the $E_z$ and $P_z$ components in the model. An incoming plane wave can be written as  $\bmE_{in}(z,\bm \rho)=\bmE_i e^{\i k_{zi}z +\i\bm\kappa_i\cdot\bm \rho}$
where $k_{zi}=\sqrt{\frac{\omega_i^2n_i^2}{c^2}-\kappa_i^2}$ with propagation condition $\omega_in_i>c\kappa_i$ and refractive index $n_i$.

The geometric anisotropy of the nanopillars can be taken into account by replacing the scalar susceptibility  $\chi_0$ by the diagonal 2x2 susceptibility tensor.  The tensor components along the x and y axes are given by $\chi_x$ and $\chi_y$, respectively. Therefore, for the rectangular nanopillar oriented along $x$ and $y$, the transmission matrix in momentum space is given by $\tilde{\bmE}=\hat{\tilde{T}} \bmE_i$, where the transmitted electric field in the momentum space is $\tilde{\bmE}=\{\tilde{E}_x,\tilde{E}_y\}$, and the incident field is  $\tilde{\bmE}_i=\{\tilde{E}_{xi},\tilde{E}_{yi}\}$. The transmission matrix then reads (see Section S2 in SM for details)
\begin{align}\label{eq:T0}
\begin{split}
\hat{\tilde{T}}=\left(
\begin{array}{cccc}
\tilde{t}_{xx} & \tilde{t}_{xy}   \\
\\
\tilde{t}_{yx}  & \tilde{t}_{yy}  \\
\end{array}
\right),
\end{split}
\end{align}
where $\tilde{t}_{ij}$, $i,j=x,y$ defined in Eq. (S17) explicitly depends on the form-factor $f_{mn,j}$. For the element oriented along $x$ and $y$ axes, i.e. $\phi_j=0$, the corresponding form-factor $f_{mn,j}(\phi_j=0)\equiv \sin(\frac{ml_x\pi}{(2N+1)a_x})\sin(\frac{nl_y\pi}{a_y})/(mn\pi^3)$.
Since the metasurface consists of nanopillars rotated around $z$ axes with the constant incremental angle $\phi_j=\frac{-\pi j}{2N+1}$ in Fig. \ref{fig1}, the corresponding rotation matrix $R(\phi_j)$ is given by
\begin{equation}\label{eq:R0}
\begin{split}
\quad
\hat{R}(\phi_j)=\left(
\begin{array}{cccc}
\cos(\phi_j) & -\sin(\phi_j) \\
\\
\sin(\phi_j)  & \cos(\phi_j) \\
\end{array}
\right).
\end{split}
\end{equation}
According to superposition principle, the transmission matrix of the metasurface can be obtained by summing the contributions of individual nanopillars, given by $\tilde{\bmE}(\mathbf{K})=\sum_j\hat{\tilde{T}}'(\phi_j)\tilde{\bmE}_i$, where the rotation-dependent transmission matrix is given by $\hat{\tilde{T}}'(\phi_j)= \hat{R}^{\dagger}(\phi_j) \hat{\tilde{T}}\hat{R}(\phi_j)$.

Using Pauli algebra for two-component polarization basis without explicit factorization of the additional propagation phase, the rotation-dependent transmission matrix reads
\begin{align}\label{eq:EK}
2\hat{\tilde{T}}'(\phi_j)&=(\tilde{t}_{xx}+\tilde{t}_{yy})\hat{I}+\i(\tilde{t}_{xy}-\tilde{t}_{yx})\hat{\sigma}_z\notag\\
&+(\tilde{t}_{xx}-\tilde{t}_{yy}) (e^{2\i\phi_j}\hat{\sigma}_-+e^{-2\i\phi_j}\hat{\sigma}_+)\notag\\
&+i(\tilde{t}_{xy}+\tilde{t}_{yx}) (e^{2\i\phi_j}\hat{\sigma}_--e^{-2\i\phi_j}\hat{\sigma}_+).
\end{align}
Here, $\hat{\sigma}_{\pm}=(\hat{\sigma}_x\pm \i\hat{\sigma}_y)/2$ is the spin-flip operator and the extra phase term $e^{\pm 2\i \phi_j}$ can be understood as the PB phase term \cite{36,pauli_chinese,Huang}.

The transmitted field in the coordinate space (the form of $\hat{T}$ is listed in Eq. (S18)) can be consecutively written as $\bmE(z,\bm\rho)=\sum_{mnj}F_{mn,j}(z,\bm\rho)\hat{T}'(\phi_j)\bmE_i(z,\bm\rho)$, where the propagation factor is
\begin{align}\label{eq:Fmnj}
F_{mn,j}=e^{\i\psi_{mn,j}}[ e^{\i k_{z}z}\mathcal{H}(z>0) +e^{-\i k_{z}z}\mathcal{H}(z<0)]|_{\bm\kappa=\bm\kappa_i},
\end{align}
where $k_{z}=\sqrt{\omega_i^2n_{t}^2/c^2-K_{||}^2}$, $K_{||}^2=(G_{x,mn}+\kappa_{x})^2+(G_{y,mn}+\kappa_{y})^2$ with momentum vectors of incident light $\bm\kappa_i=\kappa_{xi}\bm e_x+\kappa_{yi}\bm e_y$.

\subsection{Discussion of analytical results}
Analogous to the Bragg scattering in solid crystals, constructive interference of the propagating wave on the sub-wavelength periodic structure changes the complex amplitude of the refracted and reflected waves due to cumulative scattering from different crystal planes (see Eq. (\ref{eq:Fmnj})).
The evanescent waves emerge when $n\neq0$, whose momentum vectors satisfy $\omega_i^2n_{t}^2/c^2-K_{x}^2-K_{y}^2<0$. Overall, the transmitted field in zeroth diffraction order $n=0$ contains both effects of propagation and topological phases. The first line in Eq. (\ref{eq:EK}) corresponding to the co-polarization component transmitted field contains only the propagation phase $e^{\i\psi_{mn,j}}$ embedded in the propagation factor $F_{mn,j}$. The second and the third lines in Eq. (\ref{eq:EK}) yield the cross-polarization components which depend on both propagation and PB phases via $e^{\i\psi_{mn,j}\pm 2\i\phi_j}$. Due to the translation invariance, the PB phase of the individual nanopillars is distributed uniformly between $0$ and $2\pi$ such that  $\sum_j e^{\pm\i\Xi_j} \simeq 0$ except $m=0,\pm 1$, where $\Xi_j=-\bmG_{mn}\cdot j\bm a_1\pm 2\phi_j$ is the total phase. For $m=0$, only the PB phase-independent co-polarized component can be observed. By calculating the $x$-dependent part of propagation phase $\psi_{mn,j}$, we can find this component corresponds to the conventional diffraction which follows the ordinary Snell's law, $n_{t} \mathrm{sin}(\theta')-n_{i}\mathrm{sin}(\theta) =0$  where $\theta$ and $\theta'$ are the incident and transmitted angles, respectively. For $m=\pm1$, the PB phase cancels the $j$-dependent part of the propagation phase and only cross-polarized components are detectable since $G_{x,\pm10 }a_1j\equiv \mp 2\phi_j$. The remaining $x$-dependent propagation phase given by $(\frac{\pm2\pi}{(2N+1)a_1}+\kappa_{xi})x$ yields the generalized Snell's law which governs the anomalous refraction. For the light beam propagating in the $xz$ plane, the refraction angle is defined by $\mathrm{sin}(\theta')=\frac{c}{\omega_in_{t}}\left(\frac{\pm 2\pi}{(2N+1)a_1}+\kappa_{xi}\right)$. Thus for $x=[0, (2N+1)a_1]$, one can obtain
\begin{equation}\label{eq:Snell}
\mathrm{sin}(\theta') n_{t}-\mathrm{sin}(\theta)n_{i}=\frac{\pm \lambda}{(2N+1)a_1}.
\end{equation}

We now calculate the Fresnel coefficient and analyze the chiral transmission properties in the circular polarization (CP) basis: $\sigma_{\pm }=(\bm{e}_x \cos(\theta') \pm  \i\bm{e}_y)/\sqrt{2}$ where $\theta'$ is the refraction angle. Following the derivation shown in Section S3 and assuming the amplitude of the incident light is $\bmE^{s}=\cos\theta\bm{e}_x+s\i \bm{e}_y$ ($s=\pm1$), the transmitted light can be written as $\bmE=\sum_{mnj}(\bmE_1+\bmE_2e^{-s\i 2\phi_j}+\bmE_3e^{s\i 2\phi_j})F_{mn,j}$, where amplitudes $\bmE_1$, $\bmE_2$, and $\bmE_3$ are given by
\begin{equation}\label{eq:poltrans1}
\begin{split}
&\bmE_{1}=t_{1+}\bmE^{s}+t_{1-}\bmE^{-s},\\
&\bmE_{2}=t_{2-+}\bmE^{s}+t_{2++}\bmE^{-s},\\
&\bmE_{3}=t_{2+-}\bmE^{s}+t_{2--}\bmE^{-s}.\\
\end{split}
\end{equation}
Here, the coefficients $t_{1\pm}$ and $t_{2\pm\pm}$ are given in Eq. (S25).
The additional phase factor $e^{\pm s\i 2\phi_j}$, the PB phase, originates not only from the geometric rotation of the nanopillar in the unit cell, but also from the polarization of light.
In other words, the additional phase $\pm 2s\phi_j$ relies on the symmetry relation between the polarization of light and geometric nanostructures of metasurface rather than the specific coordinate system, which is a characteristic of topological phase. In order to make it more clear and easy to compare with existing research results, we discuss and summarize the selective transmission of cross-polarized beam for all possible chiral combinations of the input polarization and metasurface in Tab. 1.
For the metasurface with clockwisely rotating nanopillars depicted in Fig. \ref{fig1}B, right CP (RCP) incident light splits into  RCP light and left CP (LCP) light, while LCP incident light splits into  LCP light and RCP light. Furthermore, as we demonstrated here, the PB phase term $e^{- s\i 2\phi_j}$ of the output amplitude contributes to the effect of non-zero cross-polarized beam. If the entire metasurface is rotated counter-clockwise by $\pi$ along $z$ axis which means $\phi_j=\frac{\pi}{2N+1}$, the constant phase term is written as $e^{2\i m\phi_j}$. Then the phase gradient of cross-polarized beam changes its sign. It is clear that the sign of the phase gradient is determined by the handedness of incident light and metasurface.
\begin{table}[htb]
 \centering
	\caption{\label{tab:table1}Cross-polarized transmission for different combinations of the input polarization and metasurface }
	\begin{tabular}{l|ccc}
\hline
\hline
	Antenna rotation & Input & Output(order) & Phase gradient\\
	\hline
Clockwise & $\sigma_+$ & $\sigma_-$ (+1) & $\frac{ \lambda}{(2N+1)a_1}$\\
\cline{2-4}
	  & $\sigma_-$ & $\sigma_+$ (-1) & $\frac{ -\lambda}{(2N+1)a_1}$\\
\cline{2-4}
      & $LP$ & $\sigma_-$ (+1) & $\frac{ \lambda}{(2N+1)a_1}$\\
      &  & $\sigma_+$ (-1) & $\frac{ -\lambda}{(2N+1)a_1}$\\
   \hline
Counter-clockwise & $\sigma_+$ & $\sigma_-$ (-1) & $\frac{ -\lambda}{(2N+1)a_1}$\\
\cline{2-4}
			  & $\sigma_-$ & $\sigma_+$ (+1) & $\frac{ \lambda}{(2N+1)a_1}$\\
\cline{2-4}
             & $LP$ & $\sigma_-$ (-1) & $\frac{ -\lambda}{(2N+1)a_1}$\\
      &  & $\sigma_+$ (+1) & $\frac{ \lambda}{(2N+1)a_1}$\\
      \hline
		\end{tabular}
\begin{tablenotes}
    \centering
	\footnotesize
	\item[1] LP denotes linear polarization.
\end{tablenotes}
\end{table}

For an arbitrary input polarization, we can decompose the normally incident light in the CP basis as  $\mathbf{E}_{in} \equiv \mathbf{E}_{||}=\alpha \sigma_+ + \beta \sigma_-$ with $\beta=\sqrt{1-\alpha^2}$ and considering a normal incidence $\theta'=0$. The transmitted light can be then recast as
\begin{equation}\label{eq:berryp}
\mathbf{E}=\sum_{j,m=\pm1}(
t_{||}F_{00,j}\mathbf{E}_{||}
+t_{\bot}F_{m0,j}
M(\phi_j)\mathbf{E}_{\bot}),
\end{equation}
where $t_{||}=\frac{t_{xx} + t_{yy}}{2}$, $t_{\bot}=\frac{t_{xx} - t_{yy}}{2}$, $\mathbf{E}_{||} \cdot \mathbf{E}_{\bot}=0$, and $M(\phi_j)=\left(
\begin{array}{cccc}
e^{\i 2\phi_j} & 0 \\
0 & -e^{-\i 2\phi_j}\\
\end{array}
\right)$.
The corresponding transmission of the co- and cross-polarized beams for an arbitrary polarization incident light are illustrated schematically in Fig. \ref{fig1}A. Depending on the combination of the incident polarization and geometric rotation of the nanopillar, a cross-polarized retardation with a positive or negative phase occurs, leading to self-constructive or self-destructive interference effects. Figure \ref{fig1}A indicates the relative phase retardation $\delta \Phi_{\pm}$, is a function of the interface lateral displacement $\delta(x)$ between co- and cross-polarized beams.
\begin{figure}[!t]
	\centering
	\subfigure{
		\includegraphics[width=0.6\linewidth]{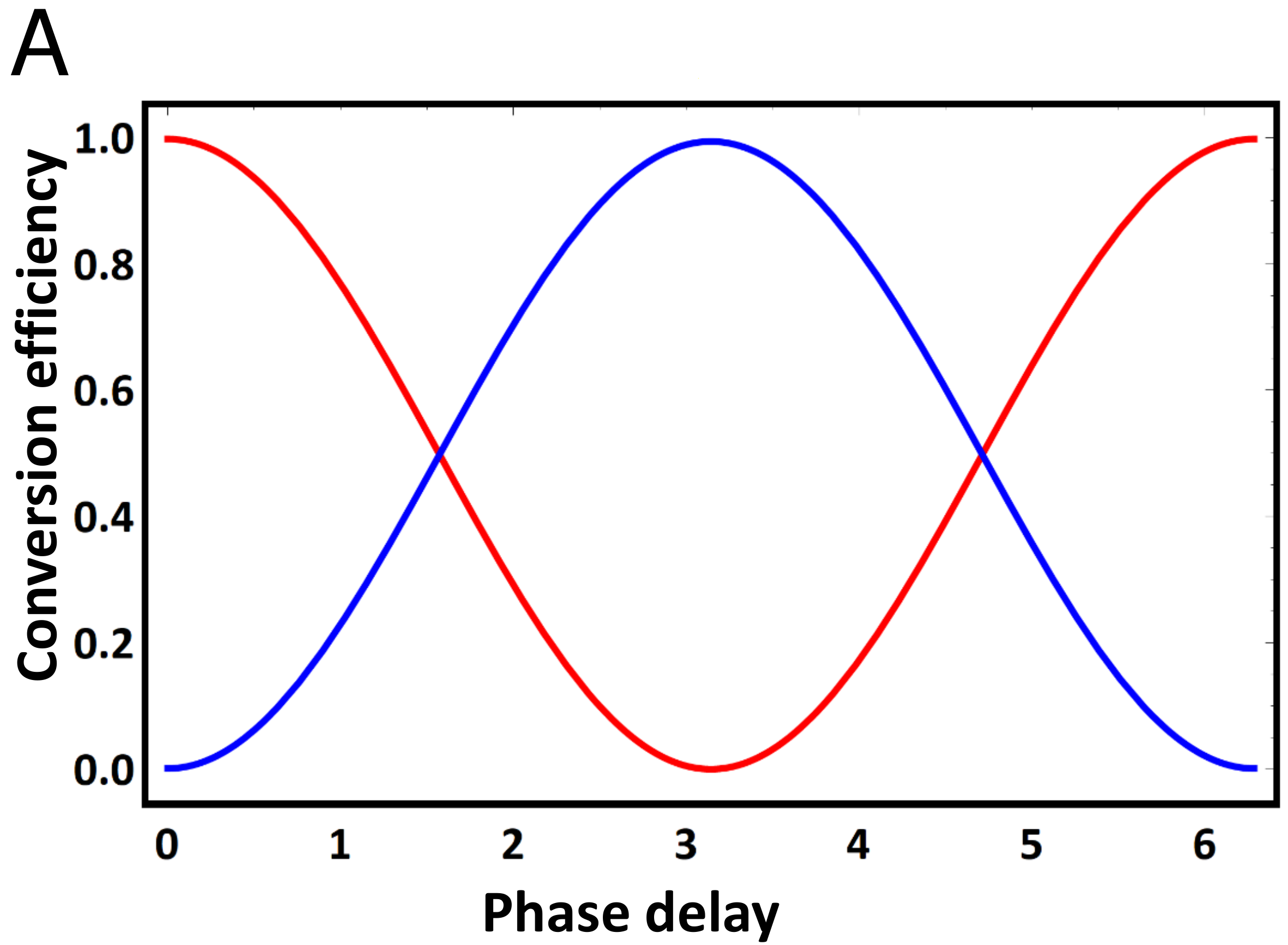}}
	\subfigure{
		\includegraphics[width=0.48\linewidth]{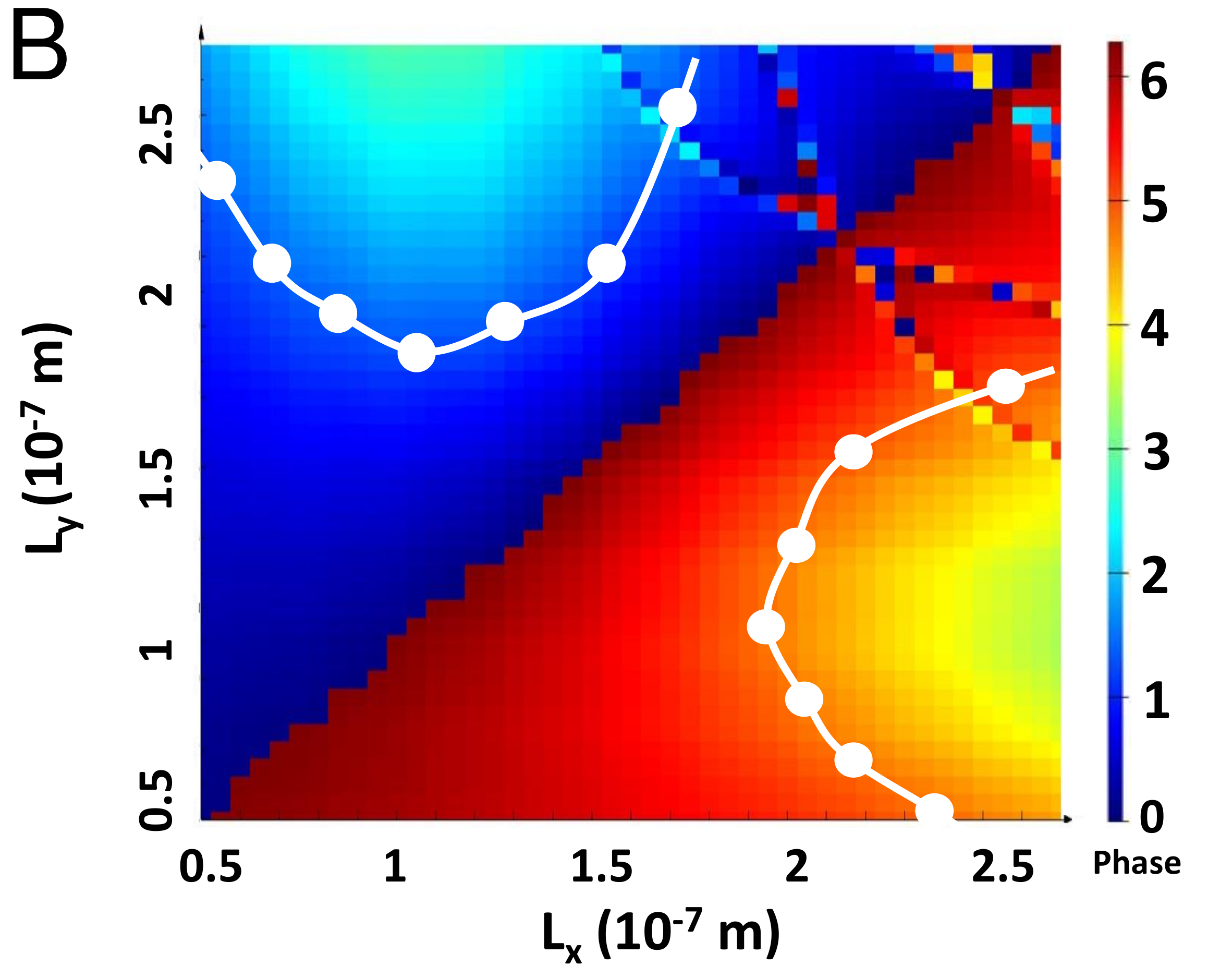}}
\subfigure{
		\includegraphics[width=0.48\linewidth]{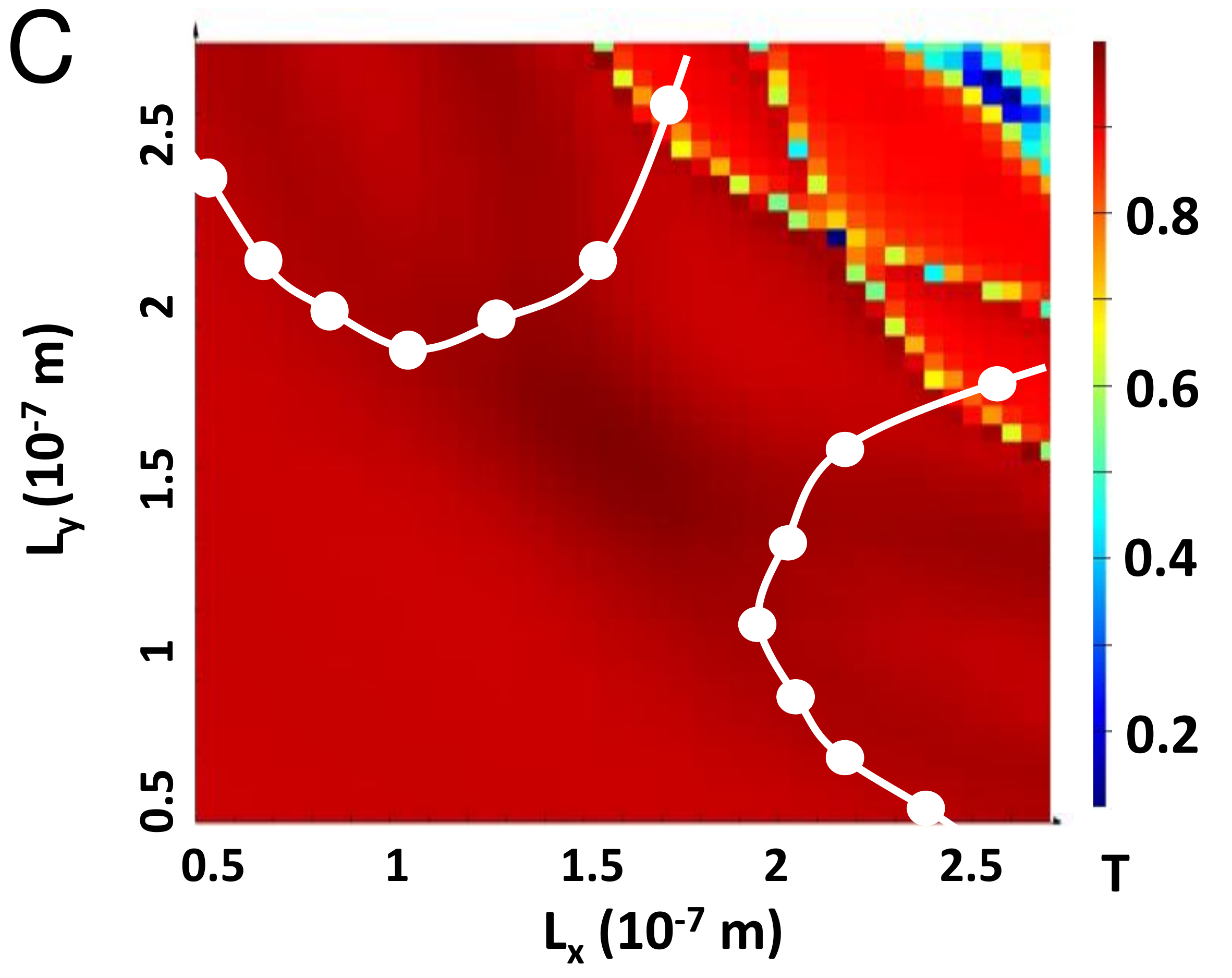}}
\subfigure{
		\includegraphics[width=0.48\linewidth]{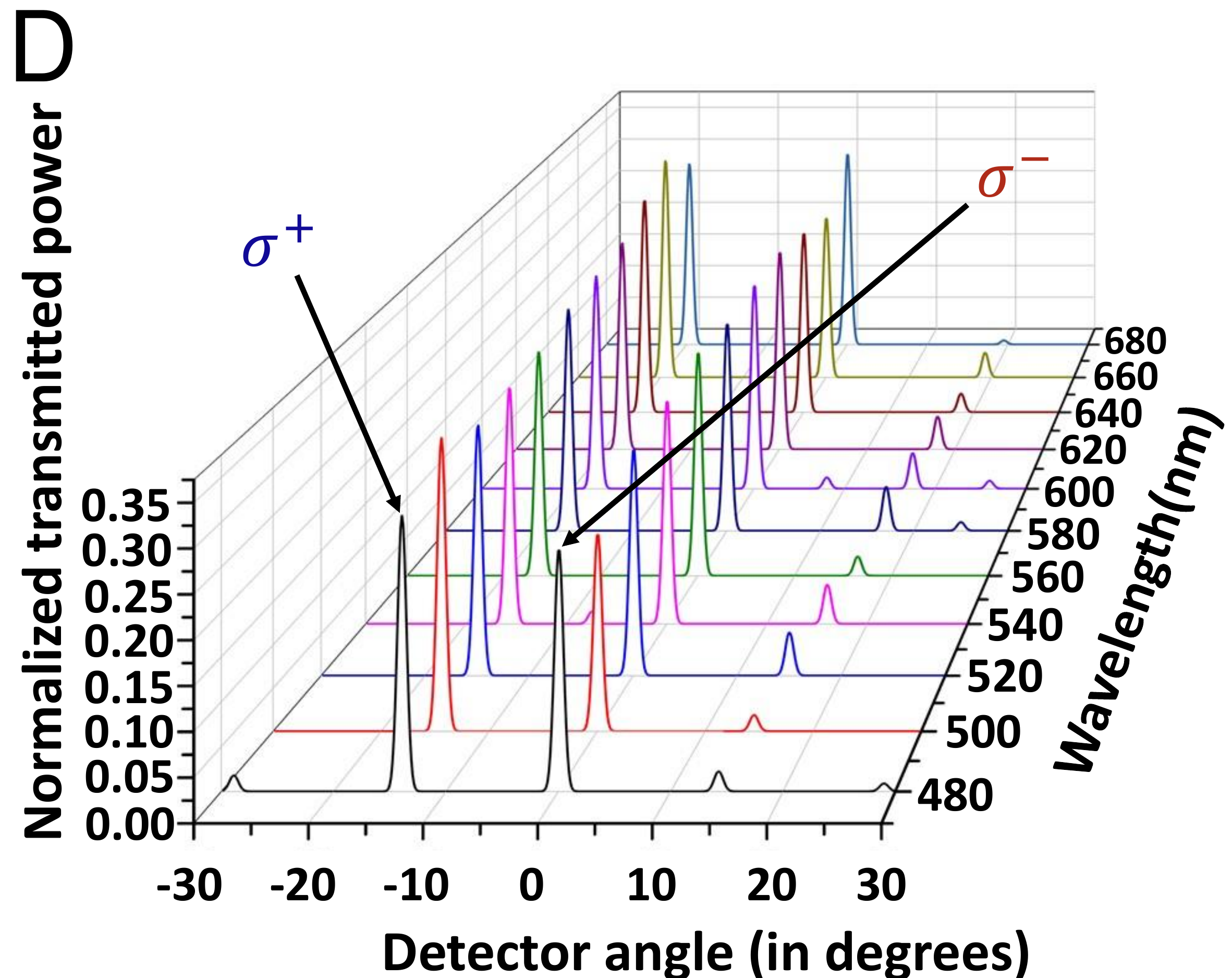}}
\subfigure{
		\includegraphics[width=0.48\linewidth]{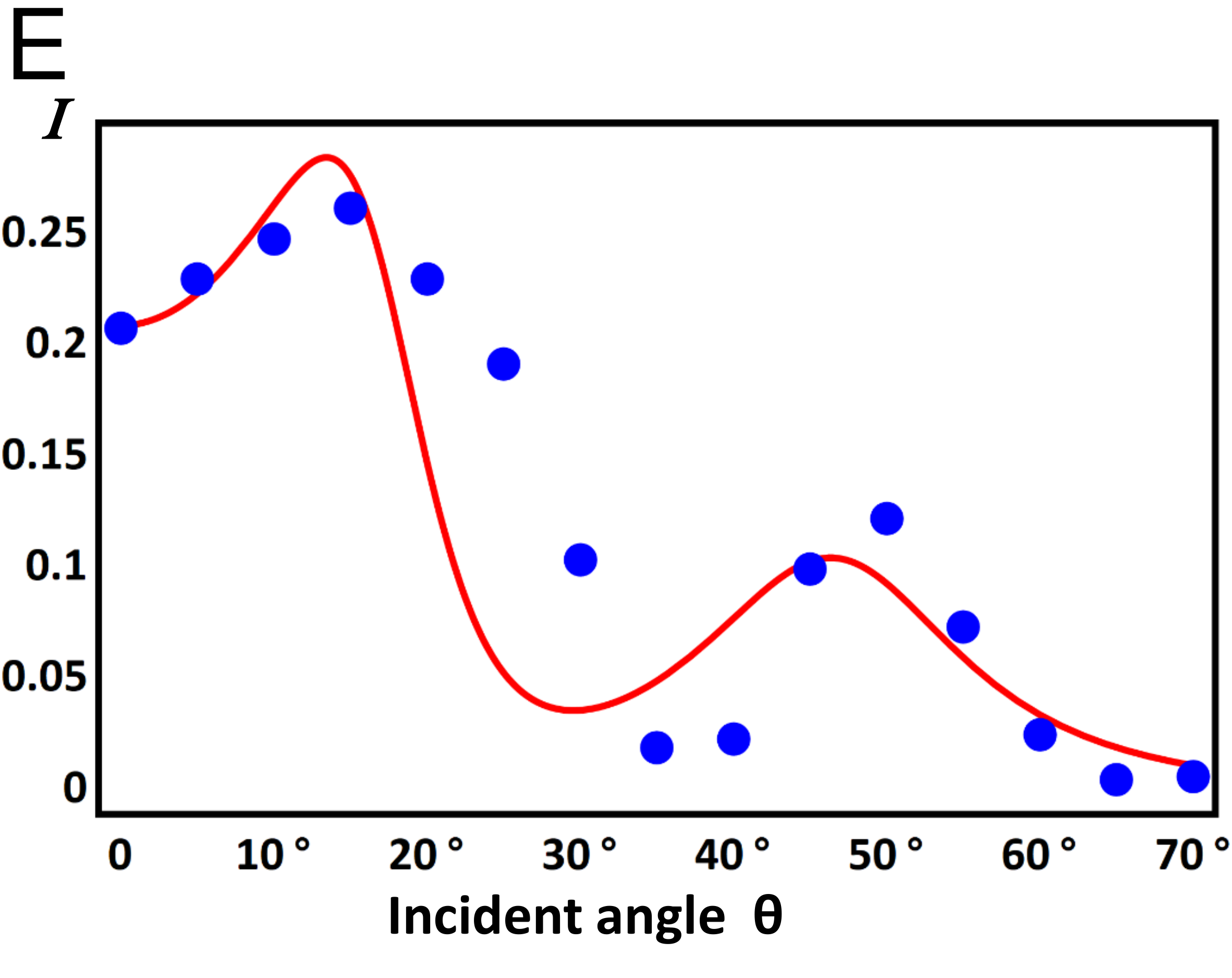}}
	\caption{\footnotesize (A) Calculated polarization conversion efficiency (blue), co-polarization transmission(red), of the subwavelength array of PB nanopillars as function of the delay between polarization eigenstates. (B) and (C) Full wave numerical simulations performed to extract the phase retardation between $E_x$ and $E_y$ components (B) and transmission maps (C) as functions of length and width of the nanopillars.  (D) Experimental measurements of the normalized transmission across a PB metasurface designed according to the guideline in (B) and (C) as a function of the incidence angle changes for LCP ($\sigma_{-}$) incidence. (E) Comparison between experiments and theory of the anomalous refraction efficiency as a function of the incident angle, where $I$ is the transmitted power. Parameters of the simulations are $a_1=500 $nm, $a_2=400 $nm, $l_x=260 $nm, $l_y=85 $nm, $l_z=632.8 $nm, $\lambda=632.8 $nm, $n_i=1.61+0.3\i$, $n_t=1.2-0.001\i$, and $\chi_{x,y}$ (see Eq. (S8) in SM) with $\omega_0=2.75$ PHz, $\omega_1=1.71$ PHz, and $n_{\text{eff}}=1.2-0.01\i$ account for the Fresnel coefficient at the first interface (see Section S3.1 in SM for details).}
	\label{fig2}
\end{figure}

\section{Interferometric measurement of the topological phase}

Therefore, the PB phase results in the opposite phase delays on the orthogonal CP components.  The relevant phenomena, such as generalized Snell's law, arbitrary polarization holography \cite{7,36}, optical edge detection \cite{zhou2019optical} and the photonic spin Hall effect\cite{zentgraf_2019, Zhang_2013}, can be thus described using our theory.. In the following, we focus on topological phase characterization using the polarization-dependent translational symmetry breaking measurement based on the Mach-Zehnder interferometer (MZI). The GaN-based PB metasurface is used as a $50/50$ CP beam splitter in the performance of self-phase referencing. To better understand the design of the birefringent nanostructure, we theoretically calculate the co- and cross-polarized scattering amplitudes of an array of identical nanopillars as a function of the phase delay between $x$ and $y$ polarization, i.e. tuning the phase difference of the diagonal elements of susceptibility tensor which represents the geometric anisotropy of the metasurface. As shown in Fig. \ref{fig2}A, the ratio of the co- and cross-polarized transmission amplitude reach $50/50$ when the the phase difference of the diagonal elements of susceptibility tensor is $\pi/2$ or $3\pi/2$.
In order to identify GaN nanopillars with $\pi/2$ or  $3\pi/2$ phase delay between $x$ and $y$ polarizations, full wave numerical simulations is performed to extract the phase retardation between $E_x$ and $E_y$ components and also the transmission efficiency as function of length and width of the nanopillars in Fig 2B and C. The white lines indicate the regions for which the phase delay between $x$ and $y$ polarizations is equal to $\pi/2$ and $3\pi/2$, needed to adjust amplitudes for the interferometric characterization of the PB phase. According to these theoretical prediction,
dimensions of GaN nanopillars used were length $l_x = 260$ nm, $l_y = 85$ nm and height $800$ nm. These dimensions generate phase retardation $3\pi/4$ between Ex and Ey components(see Section S4 for more details). We create the arrays of rotated nanopillars, each rotated by an angle $\pi/5$ from its neighboring element as indicated in Fig. \ref{fig1}. The whole metasurface is of the size $250\mu m$X$250\mu m$ array. The nanofabrication of metasurface was realized by patterning a $800$ nm thick GaN thin film grown on a double side polished c-plan sapphire substrate via a Molecular Beam Epitaxy (MBE) RIBER system. The GaN nanopillars were fabricated using a conventional electron beam lithography system (Raith ElphyPlus, Zeiss Supra $40$) process with metallic Nickel (Ni) hard masks through a lift-off process. To this purpose, a double layer of around $200$ nm Poly(methyl methacrylate) (PMMA) resists (495A4 then 950A2) was spin-coated on the GaN thin film, prior to baking the resist at a temperature of 125 \textcelsius. E-beam resist exposition was performed at $20$ keV. Resist development was realized with $3:1$ Isopropyl Alcohol (IPA):Methyl isobutyl ketone (MIBK) and a 50 nm thick Ni mask was deposited using E-beam evaporation. After the lift-off process in the acetone solution for 4 hours, GaN nanopillar patterns were created using reactive ion etching (RIE, Oxford system) with a plasma composed of \ce{Cl2CH4Ar} gases. Finally, the Ni mask on the top of GaN nanopillars was removed by using chemical etching with $1:2$ solution of \ce{HCl}:\ce{HNO3}.

Three gratings were designed and fabricated with different periodic arrangements of rotated nanopillars with periods $2$, $2.9$ and $4$ $\mu$m, respectively. The refraction properties of these designed metasurfaces are measured as the experimental verification of theoretically predicted 50/50 PB metasurface beam splitter. The measurements have been realized using a conventional diffraction setup, comprizing a Si-detector plugged into a lock-in amplifier to improve the dection signal to noise ratio. Acquiring the refracted signal as a function of the transmission angle, the detector rotates in a circular motion from $-30$ to $30$ degrees.
Spectral refraction response was obtained by sweeping the wavelength of a supercontinuum source coupled to a tunable single line filter in the range of $480-680$ nm, by intervals of $20$ nm.
A linear polarizer followed by a quarter waveplate was utilized to select the state of the incident polarization. As shown in Fig. \ref{fig2}D (Fig. S5), the designed metasurface can stably realize the function of 50/50 beam splitter in the wavelength range of $480-680$ nm.
For normal LCP incident light, the zeroth order occurs at 0 degrees. Both diffracted -1st (dominant) and  1st orders (weak residual signals at opposite refraction angle) are a consequence of the PB phase gradient.
The amplitudes of these two dominant co-CP and cross-CP remain $50/50$ when the incident wavelength changes as shown in Fig. \ref{fig2}D. As shown in Fig. \ref{fig2}E, the experimentally measured transmission efficiency of cross-polarized beam has two well-resolved peaks around $\ang{15}$ and $\ang{48}$ which is in agreement with analytically predicted diffraction efficiency (red curve).
\begin{figure}[!t]
	\centering
	\subfigure{
	\includegraphics[width=0.85\linewidth]{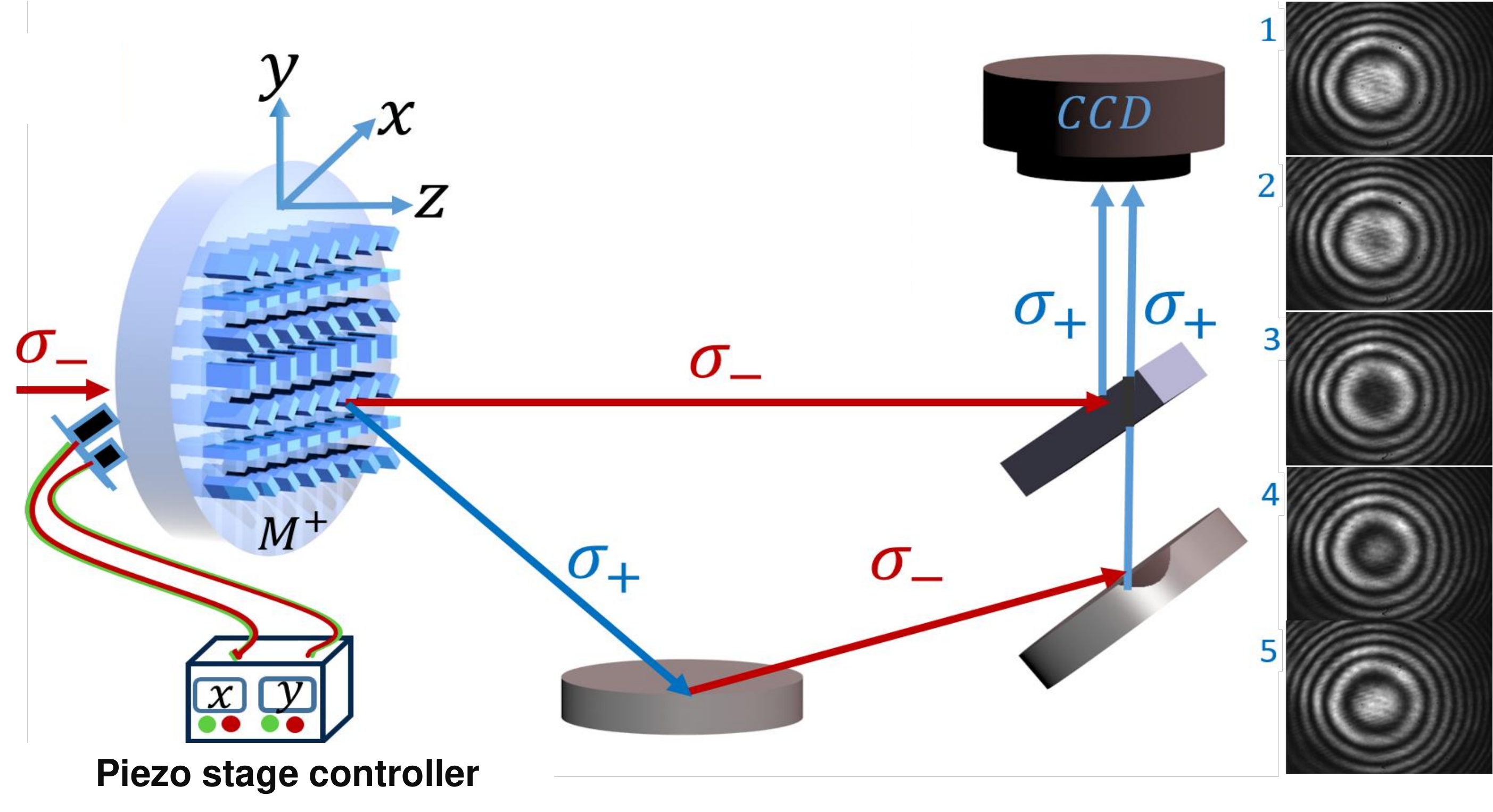}}
	\subfigure{
	\includegraphics[width=0.85\linewidth]{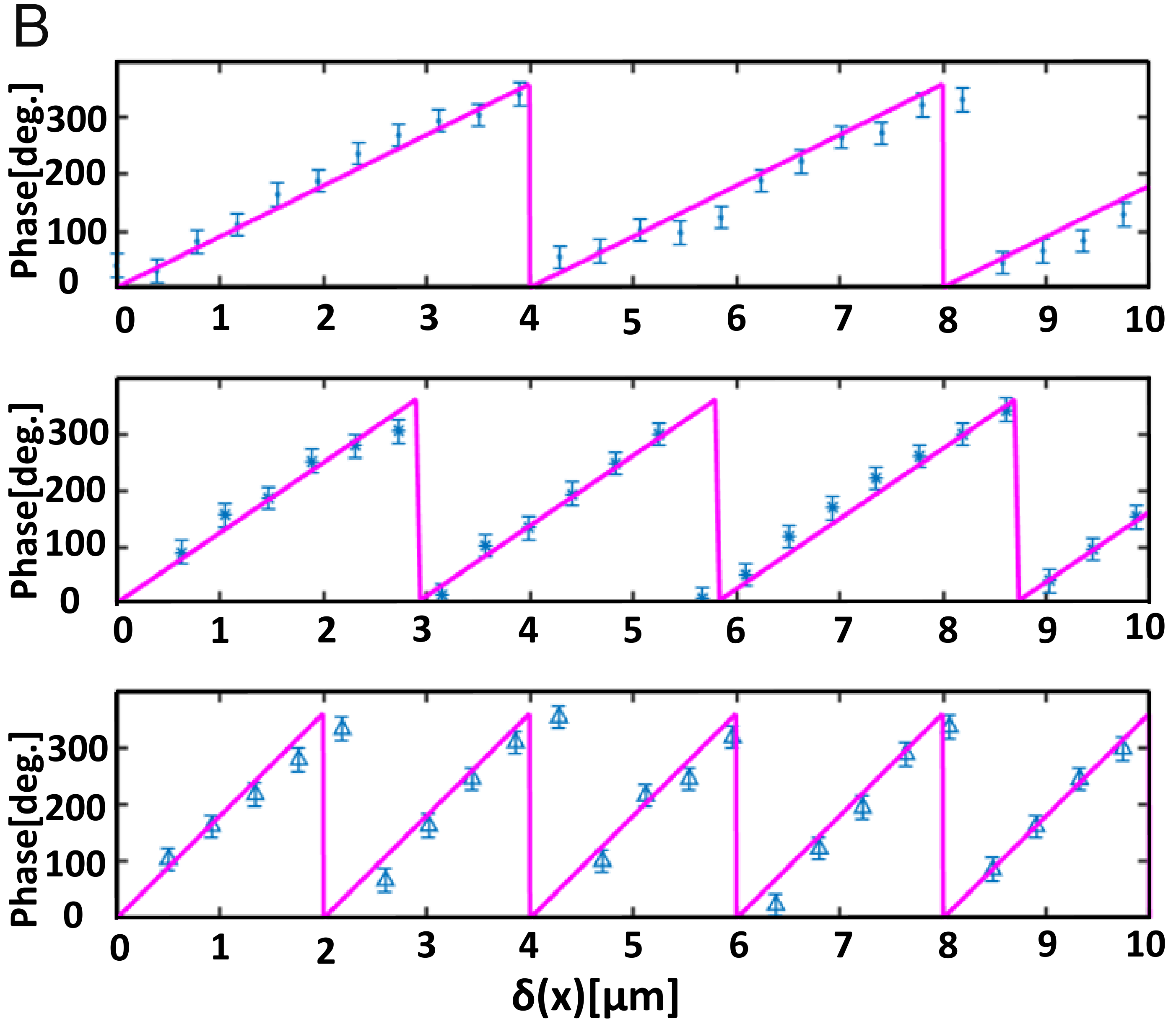}}
	\caption{\footnotesize (A) (Left) Schematic of the interferometric measurement for the characterization of the topological phase shift introduced by PB metasurface as a $50/50$ CP beam splitter. (Right) The interference fringes displacement according to  the  phase gradient direction $\delta x$, resulting from the topological phase delay shift introduced on the anomalous beam. (B) The measured phase delays as a function of the displacements are reported for 3 different gratings, with periods $\Gamma=4,2.9$ and $2 \mu m$ from top to bottom, respectively.}
	\label{fig3}
\end{figure}

We have experimentally characterized the topological phase using a self-interferometric measurement in a MZI configuration, replacing a beam splitter by the metasurface as shown in Fig. \ref{fig3}A. Phase retardation of the anomalous refracted signal as a function of the lateral displacement of the metasurface, introduced by the shifting of metasurface along the phase gradient along the $x$ axes, is recorded by monitoring the displacement of the interferogram fringes on a CCD camera after careful recombination and adjustment of the polarization handedness. The piezo stage controller is utilized to achieve minute translation of the metasurfaces as required for phase characterization in experiments discussed in Fig. \ref{fig3}B.
In the present configuration, one arm of the MZI originates from the first order refraction from the metasurface. In addition to the anomalous refraction, the metasurface imposes a phase
\begin{align}\label{eq:Phipm}
\Phi_{\pm}(x)=G_{\pm 10,x}\delta(x),
\end{align}
which is proportional to the metasurface displacement $\delta(x)$ along the phase gradient direction $x$. We propose to experimentally measure this phase by recombining both arms on a beam splitter, and recording the resulting intensity profile as a function of the translation distance. The transmitted light of RCP/LCP incidence is $\mathbf{E}=\sum_{j}(
t_{||}\sigma_{\pm}e^{\i \omega_i n_{t}z /c }
-t_{\perp}
e^{i\Phi_{\pm}(x)}\sigma_{\mp}e^{\i \sqrt{\omega_i^2n_{t}^2/c^2-G_{\pm 10,x}^2}z})\mathcal{H}(z>0)$.
The corresponding PB-dependent intensity measurement in the setup shown in Fig. 3A reads
\begin{equation}\label{eq:poltrans}
   I^+=\frac{I_{||}+I_\perp}{2}(1+r\cos(\Phi_{\pm}(x))).
\end{equation}
Here $I_{||}=|\sum_{j}t_{||}|^2$, $I_{\perp}=|\sum_{j}t_{\perp}|^2$ are the intensity of co- and cross-polarized  transmission respectively, and $r=\frac{\sqrt{I_{||}I_\perp}}{I_{||}+I_\perp}$. Then the interference fringes displacement shown in right side of Fig. \ref{fig3}A provides indirect, yet unambiguous and conclusive measurement of the PB phase.
As shown in Fig. \ref{fig3}B, we observe the linear phase variations with the wrapping periods equal to the PB phase of the metasurface in agreement with the our theoretical result in Eq. (\ref{eq:Snell}).

\section*{Conclusion}
In summary, we provide an in-depth analysis of topological PB metasurfaces by comparing experimental results obtained with spatially oriented subwavelength birefringent nanostructures, with a mesoscopic theory. This work, which demonstrates the origin of both controllable phase retardation effects, namely the propagation phase and the PB phase, is a first step in developing an intuitive understanding of topological and functional beam splitters for future applications in quantum optics and their implementations in relevant quantum information protocols based on metasurfaces, which is an important future research direction in this field \cite{21,22,23, 24, 25,25b, 26}.

\section*{Acknowledgement}
Z.G thanks S. Jiang, P. Saurabh,  G. Zhu for valuable discussions. Z.G, V.O and K.E.D gratefully acknowledge the support from National Science Foundation of China (No. 11934011), Zijiang Endowed Young Scholar Fund, East China normal University and Overseas Expertise Introduction Project for Discipline Innovation (111 Project, B12024). KD is grateful for the support of “Fédération Doeblin". P.G, R.S, and G.B acknowledge funding from the European Research Council (ERC) under the European Union's Horizon 2020 research and innovation programme (Grant agreement no. 639109).


\begin{thebibliography}{9}

\bibitem{1}
Genevet P, Capasso F, Aieta F, Khorasaninejad M, Devlin R. Recent advances in planar optics: from plasmonic to dielectric metasurfaces. \emph{Optica} (2017);\textbf{4}:139-152.

\bibitem{2}
Chen WT, Zhu AY, Sanjeev V, et~al. A broadband achromatic metalens for focusing and imaging in the visible.
\emph{Nat Nanotechnol} (2018);\textbf{13}:220-226.

\bibitem{3}
Yu N, Genevet P, Kats MA, et~al. Light propagation with phase discontinuities: Generalized laws of reflection and refraction. \emph{Science} (2011);\textbf{334}:333-337.

\bibitem{4}
Kim I, Yoon G, Jang J, et~al. Outfitting next generation displays with optical metasurfaces.
\emph{ACS Photonics} (2018);\textbf{5}:3876-3895.

\bibitem{6}
Pors A, Nielsen MG, Eriksen RL, Bozhevolnyi SI. Broadband focusing flat mirrors based on plasmonic gradient metasurfaces. \emph{Nano Letters} (2013);\textbf{13}:829-834.

\bibitem{7}
Arbabi A, Horie Y, Bagheri M, Faraon A. Dielectric metasurfaces for complete control of phase and polarization with subwavelength spatial resolution and high transmission.
\emph{Nat Nanotechnol} \textbf{10} (2015),
  937-943.

\bibitem{8}
Khorasaninejad M, Chen WT, Devlin RC,  et~al., Metalenses at visible wavelengths: Diffraction-limited focusing and subwavelength resolution imaging.
\emph{Science}  (2016);\textbf{352}:1190-1194.

\bibitem{9}
Arbabi A, Arbabi E, Kamali SM, et~al., Miniature optical planar camera based on a wide-angle metasurface doublet corrected for monochromatic aberrations.
\emph{Nat Commun} (2016);\textbf{7}:13682.

\bibitem{10}
Chen WT, Yang K, Wang C, et~al. High-efficiency broadband meta-hologram with polarization-controlled dual images.
\emph{Nano Letters} (2014);\textbf{14}:225-230.

\bibitem{11}
Zheng G, Muhlenbernd H, Kenney M, et~al. Metasurface holograms reaching 80\% efficiency.
\emph{Nature Nanotechnol} (2015);\textbf{10}:308-312.

\bibitem{12b}
Ren H, Briere G, Fang X, et~al. Metasurface orbital angular momentum holography. \emph{Nat Commun} (2019);\textbf{10}:1-8.

\bibitem{13}
Lin J, Mueller JPB, Wang Q, et~al. Polarization-controlled tunable directional coupling of surface plasmon polaritons.
\emph{Science} (2013);\textbf{340}:331-334.

\bibitem{14}
Ding F, Chen Y, Bozhevolnyi SI, Metasurface-based polarimeters. \emph{Applied Sciences} (2018);\textbf{8}:594.

\bibitem{15b}
Rubin NA, Daversa G, Chevalier P, et~al. Matrix fourier optics enables a compact full-stokes polarization camera.
\emph{Science} (2019);\textbf{365}.

\bibitem{17}
Kamali SM, Arbabi A, Arbabi E, Horie Y, Faraon A, Decoupling optical function and geometrical form using conformal flexible dielectric metasurfaces. \emph{Nat Commun} (2016);\textbf{7}:11618-11618.

\bibitem{18}
Burch J, Wen D, Chen X, Falco AD. Conformable holographic metasurfaces.
\emph{Sci Rep}  (2017);\textbf{7}:4520.

\bibitem{19}
Burch J, Di~Falco A. Surface topology specific metasurface holograms.
\emph{ACS Photonics}  (2018);\textbf{5}:1762-1766.

\bibitem{20}
Wu K, Coquet P, Wang QJ, Genevet P. Modelling of free-form conformal metasurfaces.
\emph{Nat Commun} (2018);\textbf{9}:1-8.

\bibitem{20b}
Dehmollaian M, Chamanara N, Caloz C. Wave scattering by a cylindrical metasurface cavity of arbitrary cross section: Theory and applications.
\emph{IEEE T Antenn Propag}  (2019);\textbf{67}:4059-4072.

\bibitem{20c}
Roberts C, Inampudi S, Podolskiy V A. Diffractive interface theory: nonlocal susceptibility approach to the optics of metasurfaces. \emph{Opt Express} (2015);\textbf{23}:2764-2776.

\bibitem{20d}
Achouri K, Caloz C. Design, concepts, and applications of electromagnetic metasurfaces. \emph{Nanophotonics} (2018);\textbf{7}:1095-1116.

\bibitem{20e}
Momeni A, Rajabalipanah H, Abdolali A, Achouri K. Generalized optical signal processing based on multioperator metasurfaces synthesized by susceptibility tensors.
\emph{Phys Rev Appl} (2019);\textbf{11}:064042.

\bibitem{Khanikaev}
Khanikaev A, Arju N, Fan Z, et al. Experimental demonstration of the microscopic origin of circular dichroism in two-dimensional metamaterials. \emph{Nat Commun} (2016);\textbf{7}:12045.

\bibitem{sunet}
Sun S, He Q, Xiao S, Xu Q, Li X, Zhou L.  Gradient-index meta-surfaces as a bridge linking propagating waves and surface waves. \emph{Nat Mater}, (2012);\textbf{11}(5):426-431.

\bibitem{Vahabzadeh}
Vahabzadeh Y, Chamanara N, Caloz C, Generalized Sheet Transition Condition FDTD Simulation of Metasurface. \emph{IEEE T Antenn Propag}, (2018);\textbf{66}(1):271-280.

\bibitem{de2007colloquium}
De~Abajo FJG. Colloquium : Light scattering by particle and hole arrays. \emph{Rev Mod Phys} (2007);\textbf{79}:1267-1290.

\bibitem{cz2013enhancement}
Czaplicki R, Husu H, Siikanen R, et~al. Enhancement of second-harmonic generation from metal nanoparticles by passive elements.
\emph{Phys Rev Lett} (2013);\textbf{110}:093902.

\bibitem{Liu2017Momentum}
Liu W, Li Z, Cheng H, Chen S, Tian J. Momentum analysis for metasurfaces.
\emph{Phys Rev Appl} (2017);\textbf{8}:014012.

\bibitem{smith2011analysis}
Smith DR, Tsai Y, Larouche S. Analysis of a gradient index metamaterial blazed diffraction grating. \emph{IEEE Antenn Wirel Pr} (2011);\textbf{10}:1605-1608.

\bibitem{larouche2012reconciliation}
Larouche S, Smith DR. Reconciliation of generalized refraction with diffraction theory. \emph{Opt Lett} (2012);\textbf{37}:2391-2393.

\bibitem{36}
Mueller JPB, Rubin NA, Devlin RC, Groever B, Capasso F. Metasurface polarization optics: Independent phase control of arbitrary orthogonal states of polarization.
\emph{Phys Rev Lett} (2017); \textbf{118}:113901.

\bibitem{Pancharatnam1956}
Pancharatnam S. Generalized theory of interference, and its applications: Part I. coherent pencils.
\emph{Proc Indian Acad Sci Sect A} (1956);\textbf{44}:247-262.

\bibitem{berry1987the}
Berry MV. The adiabatic phase and pancharatnam's phase for polarized light. \emph{J Mod Optic} (1987);\textbf{34}: 1401-1407.

\bibitem{kuratsuji1998maxwellsch}
Kuratsuji H, Kakigi S. Maxwell-schrodinger equation for polarized light and evolution of the stokes parameters.
\emph{Phys Rev Lett} (1998);\textbf{80}:1888-1891.

\bibitem{bliokh2006conservation}
Bliokh KY, Bliokh YP. Conservation of angular momentum, transverse shift, and spin hall effect in reflection and refraction of an electromagnetic wave packet.
\emph{Phys Rev Lett}  (2006);\textbf{96}:073903.

\bibitem{zhu2019generalized}
Zhu T, Lou Y, Zhou Y, et~al. Generalized spatial differentiation from the spin hall effect of light and its application in image processing of edge detection. \emph{Phys Rev Appl} (2019);\textbf{11}:034043.

\bibitem{pauli_chinese}
Luo W, Xiao S, He Q, Sun S, Zhou L. Photonic spin hall effect with nearly 100\% efficiency.
\emph{Adv Opt Mater}  (2015);\textbf{3}:1102-1108.

\bibitem{Huang}
Huang L, Chen X, Muhlenbernd H, et~al. Dispersionless phase discontinuities for controlling light propagation.
\emph{Nano Lett} (2012);\textbf{12}:5750-5755.

\bibitem{Hasman_2002}
Bomzon Z, Biener G, Kleiner V, Hasman E. Space-variant Pancharatnam–Berry phase optical elements with computer-generated subwavelength gratings.
\emph{Opt Lett} (2002);\textbf{27}:1141-1143.

\bibitem{Brongersma_2014}
Lin D, Fan P, Hasman E, Brongersma ML. Dielectric gradient metasurface optical elements. \emph{Science} (2014);\textbf{345}:298-302.

\bibitem{Landau}
Landau LD, Lifshitz EM. {\em Electrodynamics of continuous media}. \newblock (Pergamon, Amsterdam), Second edition, (1984).

\bibitem{zhou2019optical}
Zhou J, Qian H, Chen C, et~al. Optical edge detection based on high-efficiency dielectric metasurface. \emph{Proc Natl Acad Sci USA} (2019);\textbf{116}:11137-11140.

\bibitem{zentgraf_2019}
Georgi P, Massaro  M, Luo K, et~al. Metasurface interferometry toward quantum sensors. \emph{Light: Sci Appl} (2019);\textbf{8}1-7.

\bibitem{Zhang_2013}
Yin X, Ye Z, Rho J, Wang Y, Zhang X. Photonic spin hall effect at metasurfaces. \emph{Science} (2013);\textbf{339}:1405-1407.

\bibitem{21}
Liberal I, Engheta N. Nonradiating and radiating modes excited by quantum emitters in open epsilon-near-zero cavities. \emph{Sci Adv} (2016);\textbf{2}.

\bibitem{22}
Sokhoyan R, Atwater HA. Quantum optical properties of a dipole emitter coupled to an $\epsilon$-near-zero nanoscale waveguide. \emph{Opt Express} (2013);\textbf{21}:32279-32290.

\bibitem{23}
Ren X, Jha PK, Wang Y, Zhang X. Nonconventional metasurfaces: From non-hermitian coupling, quantum interactions, to skin cloak. \emph{Nanophotonics} (2018);\textbf{7}:1233-1243.

\bibitem{24}
Dorfman KE, Jha PK, Voronine DV, et~al. Quantum-coherence-enhanced surface plasmon amplification by stimulated emission of radiation.
\emph{Phys Rev Lett} (2013);\textbf{111}:043601.

\bibitem{25}
Jha PK, Ni X, Wu C, Wang Y, Zhang X. Metasurface-enabled remote quantum interference. \emph{Phys Rev Lett} (2015);\textbf{115}:025501.

\bibitem{25b}
Lassalle E, Lalanne P, Aljunid S, Genevet P, Stout B, Durt T, Wilkowski D. Long-lifetime coherence in a quantum emitter induced by a metasurface. preprint (2019), https://arxiv.org/pdf/1909.02409v1.

\bibitem{26}
Stav T, Faerman A, Maguid E, et~al. Quantum entanglement of the spin and orbital angular momentum of photons using metamaterials. \emph{Science} (2018);\textbf{361}:1101-1104.

\end{thebibliography}
\end{document}


\title{Supplementary materials: Revealing topological phase in Pancharatnam-Berry metasurfaces using mesoscopic electrodynamics}

\author{Zhanjie Gao}
\email{Zhanjie Gao, and Sandeep Golla contributed equally to this work.}
\affiliation{State Key Laboratory of Precision Spectroscopy, East China Normal University, Shanghai 200062, China}
\author{Sandeep Golla}
\email{Zhanjie Gao, and Sandeep Golla contributed equally to this work.}
\affiliation{Universit\'{e} C\^{o}te d$^\prime$Azur, CNRS, CRHEA, rue Bernard Gregory, Valbonne 06560, France}
\author{Rajath Sawant}
\affiliation{Universit\'{e} C\^{o}te d$^\prime$Azur, CNRS, CRHEA, rue Bernard Gregory, Valbonne 06560, France}
\author{Vladimir Osipov}
\affiliation{State Key Laboratory of Precision Spectroscopy, East China Normal University, Shanghai 200062, China}\affiliation{Holon Institute of Technology, 52 Golomb Street, POB 305 Holon 5810201, Israel}
\author{Gauthier Briere}
\affiliation{Universit\'{e} C\^{o}te d$^\prime$Azur, CNRS, CRHEA, rue Bernard Gregory, Valbonne 06560, France}
\author{Stephane Vezian}
\affiliation{Universit\'{e} C\^{o}te d$^\prime$Azur, CNRS, CRHEA, rue Bernard Gregory, Valbonne 06560, France}
\author{Benjamin Damilano}
\affiliation{Universit\'{e} C\^{o}te d$^\prime$Azur, CNRS, CRHEA, rue Bernard Gregory, Valbonne 06560, France}
\author{Patrice Genevet}
\email{patrice.genevet@crhea.cnrs.fr}
\affiliation{Universit\'{e} C\^{o}te d$^\prime$Azur, CNRS, CRHEA, rue Bernard Gregory, Valbonne 06560, France}
\author{Konstantin E. Dorfman}
\email{dorfmank@lps.ecnu.edu.cn}
\affiliation{State Key Laboratory of Precision Spectroscopy, East China Normal University, Shanghai 200062, China}

\maketitle

\section{Electrodynamics of metasurface}
\subsection{Lattice model for light-metasurface interaction}
\begin{figure}[htb]
	\centering
	\includegraphics[width=7cm]{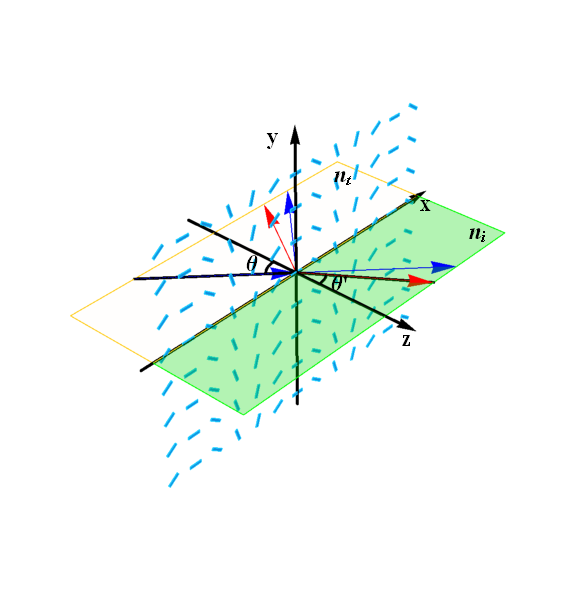}	
	\caption{\footnotesize Schematics of the reflection and refraction from the metasurface. The blue arrows represent the incidence light at the angle $\theta$,  ordinary refraction at the angle $\theta'$ and the ordinary reflection. The red arrows are the same but for the anomalous refraction and reflection. }
	\label{f2.2}
\end{figure}

The classical formulation of the light scattering problem is based on the Maxwell's equations without external currents and charges \cite{Landau}
\begin{eqnarray}\label{1.1}
\nabla\cdot\bmD &=& 0,\\
\nabla\times\bmE &=& -\frac{1}{c}\frac{\partial\bmB}{\partial t},\\
\nabla\cdot\bmB &=& 0,\\
\nabla\times\bmH &=& \frac{1}{c} \frac{\partial\bmD}{\partial t}.
\end{eqnarray}
We restrict ourself by considering non-magnetic media, so that the magnetic and diamagnetic fields are equal (the vacuum electric and magnetic permittivity are set to unity),
$\bmB=\bmH$. Following the standard Maxwell's equations notations, we use the vector of polarization, $\bmD= \bmE +4 \pi \bmP$.  Therefore the system of Maxwell's equations are given by Eq. (1) in the main text.
Information about the matter is solely contained in the functional dependence of $\bmP$ on $\bmE$. The geometry of the system is chosen such that $z$ axis of the laboratory coordinate system can be directed along the normal to the surface and $\bm \rho=(x,y)$ is a two-dimensional vector in the plane of metasurface. Particular dependence of $\bmP$ on $\bmE$ is considered in the following sections. When the metasurface is placed at the interface between two isotropic media with different refractive index, the corresponding $z$-dependent terms read
\begin{align}
\bm P(z,\bm \rho,t)=\mathcal{H}(\abs {z}\le l_z/2) \bmP(\bmE)+\mathcal{H}(z<-l_z/2)\chi_i \bmE(z,\bm \rho,t)+\mathcal{H}(z>l_z/2)\chi_t \bmE(z,\bm \rho,t).
\end{align}
Here and below Heaviside function $\mathcal{H}(\text{condition})=\set{1,\mbox{ when condition is true;~} 0, \mbox{ when condition is false}}$. Then Eq. (1) in the main text for the domain $z<-l_z/2$ and $z>l_z/2$ describe the propagating waves with the effective speeds of light $c_i=c/n_i$ and $c_t=c/n_t$ , respectively, where $n_{i,t}=\sqrt{1+4\pi\chi_{i,t}}$ denotes the refractive index of substrates on both side (see Fig. (\ref{f2.2})).

The polarization in Eq. (1) vanishes in vacuum, and has non-zero values inside the media, which represents the thin layer of metasurface. We consider a classical linear response model for the metasurface
$\bm P(z, \bm \rho,t)=\chi(\bm \rho) \bmE(z,\bm \rho,t)$.

The Fourier transformation of the fields over $\bm \rho$ are defined as follows:
\begin{eqnarray}
\bmE(z,\bm \kappa,\omega)=\int d^2\bm \rho dt  \bmE(z,\bm \rho,t) e^{-\i\bm \kappa \cdot \bm \rho+\i\omega t},\quad \bmE(\bm r,t)=\frac{1}{(2\pi)^3}\int d^2\bm \kappa d\omega \bmE(z,\bm \kappa,\omega) e^{\i\bm \kappa \cdot \bm \rho-\i\omega t}.
\end{eqnarray}
Polarization is expressed in terms of the field,
\begin{eqnarray}\label{Polariz}
\bmP(z,\bm \kappa,\omega)=\frac{1}{(2\pi)^2}\int d^2\bm \kappa'  \tilde\chi(\bm \kappa-\bm \kappa') \bmE(z,\bm \kappa',\omega), \quad \tilde \chi(\bm \kappa)= \int d^2\bm \rho  \chi(\bm \rho) e^{-\i\bm \kappa \bm \rho}.
\end{eqnarray}
The linear susceptibility function $\chi(\bm \rho)$ is a periodic function of coordinates. It can be represented as a sum of the susceptibilities of the individual primitives shifted and rotated in $xy$ plane. The reflection and transition is therefore defined by the vectors of the reciprocal lattice $\bmG$, where the function $\chi(\Delta\bm \kappa)$  reaches its maxima.


\begin{figure}[htb]
\centering
		\includegraphics[width=0.5\textwidth]{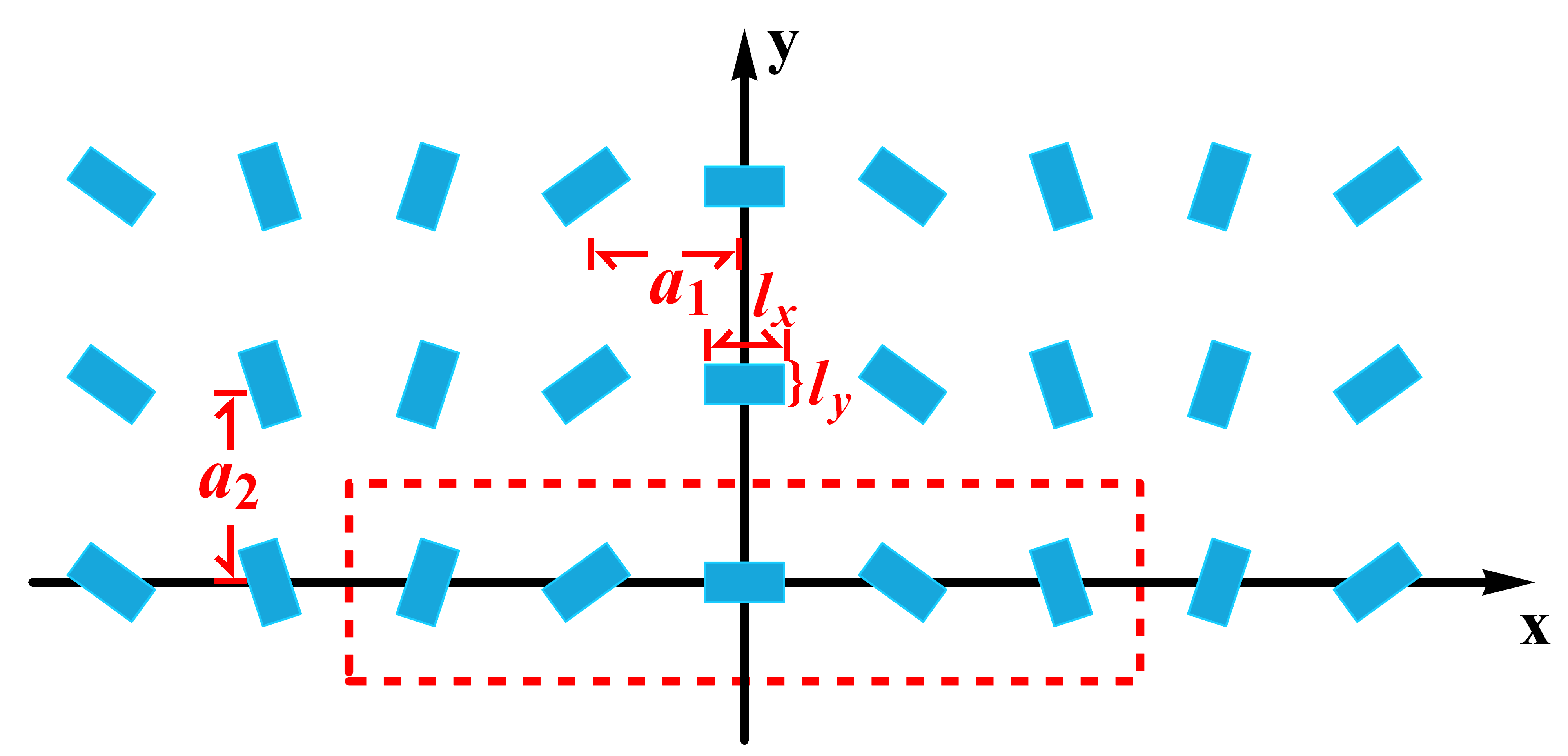}	
	\caption{\footnotesize Schematics of the metasurface. The length and width of the antenna are $l_x$ and $l_y$, respectively. The distance between the two antennas in $x$ and $y$ axis are $a_1$ and $a_2$, respectively. Along the $x$ axis, for $M=5$ (shown here) the antennas are rotated by the angle $\phi=\frac{-\pi}{(2N+1)}$.}
	\label{f2.1}
\end{figure}

\subsection{Metasurface with translational symmetry}

Since the length of the antenna is in the sub-wavelength scale and the metasurface is translation-invariant without considering the rotation of the antenna, as shown in Fig.~\ref{f2.1}, the susceptibility function $\chi(\bm \rho)$ can be represented by
\begin{equation}\label{kap}
\chi(z,\bm \rho)=\chi_0
\sum_{mn}\sum_{j=-N}^N\Omega(\bm \rho-\bm \varrho_{Mm+j,n}),
\quad
\chi_0=\left(
 \begin{array}{cccc}
  \chi_x  & 0 \\
   \\
   0  & \chi_y \\
   \end{array}
   \right).
\end{equation}
where $N=(M-1)/2$ with the number of elements in the unit cell $M$, and the susceptibilities of the nanopillar along $x$ and $y$ directions are $\chi_x=[4\pi\omega-4\pi\omega_0(1-\frac{\sin^2(\theta)}{n^2_{eff}})^{-1/2}]^{-1}$, $\chi_y=[4\pi\omega-4\pi\omega_1(1-\frac{\cos^2(\theta)}{n^2_{eff}})^{-1/2}]^{-1}$ with different resonance  frequencies $\omega_0$ and $\omega_1$ \cite{Houdre}, effective refractive index of the nanopillar $n_{eff}$ and incident angle $\theta$. The indicator function $\Omega(\bm \rho)$ describes the basic geometric primitive, which has a rectangular shape:
\begin{equation}\label{omm}
\Omega(\bm \rho)=\mathcal{H}(\abs {x}\le l_x/2)\mathcal{H}(\abs{y}\le l_y/2).
\end{equation}
The lattice translation vector $\bm \varrho_{m, n}=m \bm a_x+ n\bm a_y$ with lattice primitive translation vectors $\bm a_x=(2N+1)a_1 \bm{e}_x$ and $\bm a_y=a_2 \bm{e}_y$.
2D spatial Fourier transformation of Eq.~(\ref{kap}) is
\begin{equation}\label{eps}
\tilde \chi(\bm \kappa)=\chi_0\sum_{n}e^{-\i n\bm \kappa \bm a_2} \sum_{m}\sum_{j=-N}^N \tilde\Omega_{mnj}(\bm  \kappa)\,e^{-\i ((2N+1)m+j)\bm \kappa \bm a_1},
\end{equation}
where $\tilde\Omega_{mnj}(\bm \kappa)$ is the Fourier transform of $\Omega(\bm \rho-\bm \varrho_{Mm+j,n})$. Since the rotation of antenna in one unite cell is treated by the rotation of the transmission matrix later by the form $\hat{\tilde{T}}'(\phi_j)= \hat{R}^{\dagger}(\phi_j) \hat{\tilde{T}}\hat{R}(\phi_j)$ in the main text, all the $\tilde\Omega_{mnj}(\bm \kappa)$ with different indexes $m, n, j$ are same and the subscripts mnj will be omitted  below..
By using the Poisson summation formula $\sum\limits_{n'=-\infty}^\infty F(k-2\pi n' a)=\frac{1}{2\pi a} \sum\limits_{n=-\infty}^\infty \hat{F}(\frac{n}{2 \pi a}) e^{-\i k n /a}$ where $\hat{F}$ is the Fourier transformation of $F$, Eq. (\ref{eps}) reduce to

\begin{equation}\label{kapClass}
\tilde \chi(\bm \kappa)=  \sum_{m',n'}\delta\left(\kappa_x-\frac{2\pi m'}{(2N+1)a_1}\right)\delta\left(\kappa_y-\frac{2\pi n'}{a_2}\right)  F_{m'n'},
\end{equation}
where $F_{m'n'}=\frac{\chi_0(2\pi)^2}{(2N+1)a_1a_2} \sum\limits_{j=-N}^N \tilde\Omega(\bm  \kappa )\,e^{-\i j\bm \kappa \bm a_1}$.
Taking the inverse Fourier transform, we obtain
\begin{equation*}\label{kapClassr}
\tilde \chi(\bm \rho)= \frac{\chi_0}{(2N+1)a_1a_2} \sum_{m',n'} \sum_{j=-N}^N \tilde\Omega(\bm  G_{m'n'} )
e^{\i \bm{G}_{m'n'}\cdot( \bm{\rho}-j\bm a_1)},
\end{equation*}
where the reciprocal lattice vector $\bm{G}_{m'n'}=\frac{2\pi m'}{(2N+1)a_1}\bm{e}_x+ \frac{2\pi n'}{a_2}\bm{e}_y$.
Substitution of the result~(\ref{kapClass}) into~(\ref{Polariz}) produces the series for the polarization vector
\begin{equation}\label{pk}
\bmP(z,\bm \kappa,\omega)=
\frac{\chi_0 }{(2N+1)a_1a_2}\sum_{m',n'}
\sum\limits_{j=-N}^N\int d^2\bm \kappa'
  \delta\left(\bm \kappa-\bm \kappa'-\bm G_{m',n'}\right)
 \,e^{-\i j(\bm \kappa-\bm \kappa') \bm a_1}
\tilde\Omega(\bm \kappa-\bm \kappa')
 \bmE(z,\bm \kappa',\omega).
\end{equation}

By taking inverse Fourier transform of Eq. (\ref{pk}), polarization vector is given by Eq. (2) in the main text as a sum over the Brillouin zones.
To define the nonuniform part of the system of Maxwell's equations, we assume that the incident light has the form
$\bmE(z,\bm \rho,t)=\bmE_i \, e^{\i k_{zi}z +\i\bm\kappa_i\bm \rho-\i\omega_i t}$
where $k_{zi}=\sqrt{\frac{\omega_i^2n_i^2}{c^2}-\kappa_i^2}$ with the conditions $\omega_in_i>c\kappa_i$.
Since there are no other temporal characteristics except the $\omega_i$, all the time-derivatives can be replaced by the multiplication by $-\i\omega_i$. We seek for the solution of the Maxwell's equations (1) in the form of Eq. (4) in the main text.

\subsection{Thin metasurface limit} In zeroth approximation, we assume that the surface polarization is caused by the incident light only, so that
$
\bmP_{||}=\chi \bm E_{in}e^{\i k_{zi}+\i\bm\kappa_i\bm \rho}
$ and

\begin{equation}
\frac{\partial^2}{\partial z^2}\bmE_{||} + \nabla_{\bm\rho}^2\bmE_{||}+ \frac{\omega_i^2n_{i,t}^2}{c^2}\bmE_{||} =-4\pi\nabla_{\bm\rho}(\nabla_{\bm\rho}\cdot\bmP_{||} )-4\pi\nabla_{\bm\rho}\frac{\partial}{\partial z}P_z   - 4\pi \frac{\omega_i^2}{c^2}\bmP_{||}.
\end{equation}
The following equations are written for the $e^{\i (\bmG_{mn}+\bm{\kappa})\bm \rho}$ components of above equations. All the $\bmG_{mn}+\bm{\kappa}$ spatial components of the  $\tilde{E}, \tilde{P}$ have the same subscripts $m$ and $n$ which will be further omitted. The corresponding thin polarization component $\mathcal{H}(\abs {z}\le l_z/2)\bm P(\bm E)$ is
\begin{equation}
\begin{split}
\tilde{\mathcal{H}}(\mathrm{k}_z)\bmP_{||}(\mathrm{k}_z,\bm\kappa)=\frac{\sin(\xi l_z)}{4\pi\xi} \chi_0 f_{mn}\bmE_{||,i}\delta(\bm\kappa_i-\bm\kappa)
\simeq\frac{l_z}{4\pi} \chi_0 f_{mn}\bmE_{||,i}\delta(\bm\kappa_i-\bm\kappa),
\end{split}
\end{equation}
where $f_{m,n}=\sum_{j=-N}^N f_{mn,j} e^{-\i j\bm G_{mn} \bm a_1}$,  $2\xi=k_{zi}-\mathrm{k}_z$, $\bmE_{||,i}$ is the $x,y$ amplitudes of incident light. When the incident angle or the thickness of the antenna $l_z$ is small which corresponds to $\xi \approx 0$ or $l_z \approx 0$, then $\frac{\sin(\xi l_z)}{\xi}\approx l_z$. Considering the thickness of the metasurface to be small, we neglect the $E_z$ and $P_z$ components in the model and obtain
\begin{equation}\label{eq:Ex1}
\frac{1}{2\pi}(\frac{\omega_i^2n_{i,t}^2}{c^2}-K_{||}^2-\mathrm{k}_z^2)\tilde E_{x}(\mathrm{k}_z,\bm\kappa) = (K_x^2-\frac{\omega_i^2 }{c^2})\tilde P_{x}+ K_x K_y\tilde P_{y},
\end{equation}
\begin{equation}\label{eq:Ey1}
\frac{1}{2\pi}(\frac{\omega_i^2n_{i,t}^2}{c^2}-K_{||}^2-\mathrm{k}_z^2)\tilde E_{y}(\mathrm{k}_z,\bm\kappa) = (K_y^2-\frac{\omega_i^2 }{c^2})\tilde P_{y}+ K_yK_x \tilde P_{x},
\end{equation}
where $K_x=G_{mn,x}+\kappa_{x}$, $K_y=G_{mn,y}+\kappa_{y}$,  $K_{||}^2=(G_{mn,x}+\kappa_{x})^2+(G_{mn,y}+\kappa_{y})^2$, $\tilde E_{y}(\mathrm{k}_z,\bm\kappa), \tilde E_{y}(\mathrm{k}_z,\bm\kappa)$ are function of $\mathrm{k}_z,\bm\kappa$ need to be calculated, $\tilde\bmP_{||}=  l_z\chi_0 f_{mn}\bmE_{||,i}\delta(\bm\kappa_i-\bm\kappa)$.

\section{Reflection and refraction from metasurfaces with translational symmetry }

\subsection{Transmission matrix}
We can now investigate how the light is transmitted through the metasurface. In the present model, we assume that, for a typical design, individual primitives are spaced sparsely which eliminates a possibility of the inter-element interactions. In this case, one can calculate the transmission through the individual primitive and then sum over all the elements of the unit cell and the entire metasurface. Consider the rectangular primitive with perfectly aligned sides along x and y coordinates such that the long side $l_y$ is along the $y$-axis and the short side $l_x$ is along the $x$-axis.
We can thus  recast Eqs. (\ref{eq:Ex1}) - (\ref{eq:Ey1}) in terms of the transmission matrix $\hat{\tilde{T}}$ given by Eq. (4) in the main text (Due to $e^{-\i j\bm G_{mn} \bm a_1}$ is included in the propagation phase $\psi_{mn,j}$, we omitted it in here.):
\begin{equation*}
\begin{split}
\left(                 
  \begin{array}{cccc}
  \tilde E_{x,m,n} \\
   \\
   \tilde E_{y,m,n}
   \end{array}
\right)
&=\left(
\begin{array}{cccc}
\tilde{t}_{xx} & \tilde{t}_{xy}   \\
\\
\tilde{t}_{yx}  & \tilde{t}_{yy}  \\
\end{array}
\right)
\left(                 
  \begin{array}{cccc}
 E_{xi} \\
   \\
 E_{yi}
   \end{array}
\right),
\end{split}
\end{equation*}

where the elements of transmission matrix are given by
\begin{align}
    &\tilde{t}_{xx}=\frac{\tilde{C}'_r}{\tilde{C}'} (K_x^2-\frac{\omega_i^2 }{c^2})\chi_x,\notag\\
    &\tilde{t}_{xy}=\frac{\tilde{C}'_r}{\tilde{C}'}(K_x K_y)\chi_y,\notag\\
    &\tilde{t}_{yx}=\frac{\tilde{C}'_r}{\tilde{C}'} (K_xK_y )\chi_x,\notag\\
    &\tilde{t}_{yy}=\frac{\tilde{C}'_r}{\tilde{C}'}(K_y^2-\frac{\omega_i^2}{c^2})\chi_y.
\end{align}
Here $\tilde{C}'=-\mathrm{k}_z^2 + \frac{\omega_i^2n_{i,t}^2}{c^2}-K_{||}^2$, $\tilde{C}'_r= 2\pi l_z f_{mn,j} \delta(\bm\kappa_i-\bm\kappa)$, $\bm{\kappa}_i<\bm G_{1,1}=\frac{2\pi }{(2N+1)a_1}\bm{e}_x+ \frac{2\pi }{a_2}\bm{e}_y$.
In the physical space, the transmission matrix $\hat{T}$ can be written as
\begin{align}
    &t_{xx}=\frac{C'_r}{C'} (K_x^2-\frac{\omega_i^2}{c^2})\chi_x,\notag\\
    &t_{xy}=\frac{C'_r}{C'}(K_x K_y)\chi_y,\notag\\
    &t_{yx}=\frac{C'_r}{C'}(K_x K_y)\chi_x,\notag\\
    &t_{yy}=\frac{C'_r}{C'}(K_y^2-\frac{\omega_i^2}{c^2})\chi_y.
\end{align}
where $C'=\sqrt{ \frac{\omega_i^2n_{i,t}^2}{c^2}-K_{||}^2}\big{|}_{\bm\kappa=\bm\kappa_i}$, $C'_r= \i l_z f_{mn,j}/4\pi |_{\bm\kappa=\bm\kappa_i}$.

\section{Derivation of the Fresnel coefficient}

So far we have investigated the transmission properties of the metasurfaces assuming that the incident and transmitted light are polarized linearly. We now consider a circularly polarized (CP) light. We therefore have to transform the solution to the CP basis $\sigma_{\pm }=(\bm{e}_x \cos(\theta') \pm  \i\bm{e}_y)/\sqrt{2}$, where $\theta'$ is the refraction angle (see Fig. \ref{f2.2}). For an ordinary operator $\hat{A}$ in the linear polarization basis, we define an operator $\hat{\bar{A}}$ in the CP basis $\hat{A}
\frac{(x,y)\rightarrow (\sigma_{\pm})}{}
\hat{\bar{A}}$. For instance, the rotation operator $\hat{R}(\phi_j)$ defined in Eq. (6) can be recast in the circular polarization basis as an operator $\hat{\bar{R}}(\phi_j)$ defined as follows
\begin{equation}\label{C.3}
\begin{split}
\hat{\bar{R}}(\phi_j)=\left(
\begin{array}{cccc}
\frac{e^{\i \phi_j}(2-\sec\theta'-\cos\theta')+e^{-\i \phi_j}(2+\sec\theta'+\cos\theta')}{4}&
\frac{e^{\i \phi_j}(\sec\theta'-\cos\theta')+e^{-\i \phi_j}(-\sec\theta'+\cos\theta')}{4}
\\
\\
\frac{ e^{-\i \phi_j}(\sec\theta'-\cos\theta')+e^{\i \phi_j}(-\sec\theta'+\cos\theta')}{4}&
\frac{e^{-\i \phi_j}(2-\sec\theta'-\cos\theta')+e^{\i \phi_j}(2+\sec\theta'+\cos\theta')}{4}\\
\end{array}
\right).
\end{split}
\end{equation}
Similarly the transmission matrix $\hat{T}$ in Eq. (5) takes the form
\begin{equation}\label{C.3a}
\begin{split}
\hat{\bar{T}}\equiv\left(
\begin{array}{cccc}
\frac{t_{yy} + t_{xx} - \i t_{xy} \cos\theta' + \i t_{yx} \sec\theta'}{2} &
\frac{t_{yy} - t_{xx} - \i t_{xy} \cos\theta' - \i t_{yx} \sec\theta'}{2}\\
\\
\frac{t_{yy} - t_{xx} + \i t_{xy} \cos\theta' + \i t_{yx} \sec\theta'}{2} &
\frac{t_{yy} + t_{xx} + \i t_{xy} \cos\theta' - \i t_{yx} \sec\theta'}{2}\\
\end{array}
\right).
\end{split}
\end{equation}

We then consider the in plane transmission condition $t_{xy}=0, t_{yx}=0$. In the CP basis the corresponding operator reads
\begin{equation}
\begin{split}
\hat{\bar{T}}(\phi_j)=\left(
\begin{array}{cccc}
t'_{1j}&
t'_{2j}
\\
\\
t'^*_{2j}&
t'^*_{1j} \\
\end{array}
\right),
\end{split}
\end{equation}
where  $$t'_{1j}= \frac{4(t_{xx} + t_{yy})+e^{\i 2\phi_j}(t_{xx} - t_{yy})(\cos\theta'-\sec\theta')+e^{-\i 2\phi_j}(t_{xx} - t_{yy})(\sec\theta'-\cos\theta')}{8},$$
$$t'_{2j}= \frac{e^{-\i 2\phi_j}(t_{xx} - t_{yy})(2-\sec\theta'-\cos\theta')+e^{\i 2\phi_j}(t_{xx} - t_{yy})(2+\sec\theta'+\cos\theta')}{8}.$$ The CP light with the incident angle $\theta$ can be recast as
$\{\frac{1}{2}\left(\pm 1+ \sec\theta'\cos\theta\right),\frac{1}{2}\left(
\mp 1+ \sec\theta'\cos\theta\right),0\}$. The $E_{mn,j}$ component of the refracted light then reads
\begin{equation}
\begin{split}
E_{mn,j}=
\left(
\begin{array}{cccc}
E_{+1}+E_{+2}e^{-\i 2\phi_j}+E_{+3}e^{\i 2\phi_j}\\
\\
E_{-1}+E_{-2}e^{-\i 2\phi_j}+E_{-3}e^{\i 2\phi_j}\\
\end{array}
\right)F_{mn,j},
\end{split}
\end{equation}
where
\begin{equation}
\begin{split}
&E_{+1}=\frac{(t_{xx} + t_{yy})(\cos\theta\sec\theta'\pm1)}{4},
E_{+2}=\frac{(t_{xx} - t_{yy})(\cos\theta\pm1)(\sec\theta'-1)}{8},\\
&E_{+3}=\frac{(t_{xx} - t_{yy})(\cos\theta\mp1)(\sec\theta'+1)}{8},\\
&E_{-1}=\frac{(t_{xx} + t_{yy})(\cos\theta\sec\theta'\mp1)}{4},
E_{-2}=\frac{(t_{xx} - t_{yy})(\cos\theta\pm1)(\sec\theta'+1)}{8},\\
&E_{-3}=\frac{(t_{xx} - t_{yy})(\cos\theta\mp1)(\sec\theta'-1)}{8}.
\end{split}
\end{equation}
Consider the general refraction law, the refraction angle satisfies $\sin\theta'=\frac{m \lambda}{5a_1 n_{i,t}}+\frac{n_i \sin\theta}{n_{i,t}}$ where $m=0,\pm1$. Using the field amplitudes in the circular polarization basis $E^{\pm s}=(\bm{e}_x \cos(\theta') \pm s  i\bm{e}_y)/\sqrt{2}$, which yields the general result:
\begin{align}\label{4.1.1a}
E=\sum_{mnj} (E_1+E_2e^{-s\i 2\phi_j}+E_3e^{s\i 2\phi_j})F_{mn,j}
\end{align}
with $E_1, E_2, E_3$ can be written as Eq. (9)
where
 \begin{align}\label{eq:tpm}
&t_{1\pm}=\frac{(t_{xx} + t_{yy})(\cos\theta\sec\theta'\pm1)}{4}\notag,\\
&t_{2\pm\pm}=\frac{(t_{xx} - t_{yy})(\cos\theta\pm1)(\sec\theta'\pm1)}{8}\notag,\\
&E^{\pm s}=(\bm{e}_x \cos(\theta') \pm s  \i\bm{e}_y)/\sqrt{2}.
\end{align}
 Since $l_y << \lambda$ and $l_x<\lambda$, then $t_{yy}\propto\chi_y$ can be neglected. In order to get a more clear and simper analytic expression without effect physical picture, we discuss a 1D model in the following. Consider $t_{xx}=\frac{\i f_{mn,j}l_z \chi_x (K_x^2-\frac{\omega_i^2 }{c^2}) }{2 \sqrt{\frac{\omega_i^2n_{i,t}^2}{c^2}-K_{x}^2}}\approx t_{xx}' \cos\theta'$ where $t_{xx}'=\frac{-\i f_{mn,j} l_z \chi_x}{2 }$ and the refractive indexes in two sides of the metafursace are close $n_i=n_t\approx 1$, then $E_1, E_2, E_3$ can be written as
\begin{equation}\label{eq:poltrans1}
\begin{split}
&E_{1}=t'_{1+}E^{s}+t'_{1-}E^{-s},\\
&E_{2}=t'_{2-+}E^{s}+t'_{2++}E^{-s},\\
&E_{3}=-t'_{2+-}E^{s}-t'_{2--}E^{-s},\\
\end{split}
\end{equation}
where \begin{align}
&t'_{1\pm}=\frac{1}{4}t_{xx}'(\cos\theta\pm\cos\theta')\notag,\\
&t'_{2\pm\pm}=\frac{1}{8}t_{xx}'(\cos\theta\pm1)(\cos\theta'\pm1)\notag.\\
\end{align}
It is clear that the amplitude of different components satisfy $|t'_{1+}| >> |t'_{2-+}|,|t'_{2+-}|$ and $|t'_{2++}| >> |t'_{1-}|,|t'_{2--}|$,  when the incident and refracted angles are small. The transmission light is in the form of
 $$
 E=\sum_{mnj} (t'_{1+}E^{s}+t'_{2++}E^{-s}e^{-s\i 2\phi_j})F_{mn,j}.
 $$
 Combining with the constant phase term $e^{2\i m\phi_j}$ in $F_{mn,j}(z,\mathbf{\rho})$ (antennas are rotated in a clockwise direction in  Fig. \ref{f2.1}), for RCP  incident light ($s=1$), only $m=0, 1$ order components can be refracted;  for LCP incident light ($s=-1$), only $m=0, -1$ order components can be transferred due to the averaging out of the phase components $\sum_j e^{\pm\i 2\phi_j} \simeq \sum_j e^{\pm\i 4\phi_j} \simeq 0$.

\subsection{Derivation of the transmission efficiency}

\begin{figure}[t]
		\centering
		\includegraphics[scale=0.5]{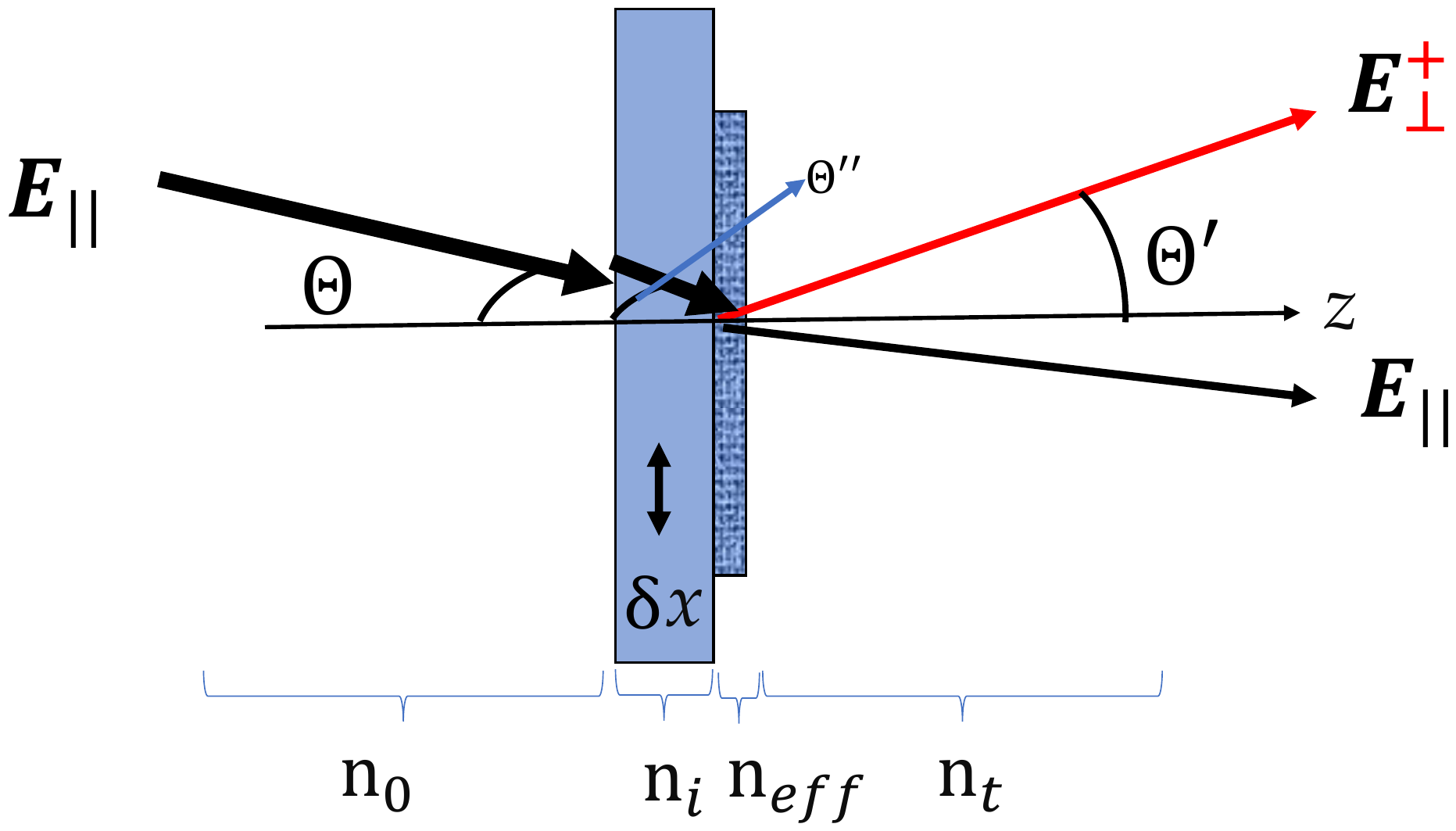}
		\caption{\footnotesize Schematic of the light transmission through
			the PB metasurface.}
		\label{figm1}
	\end{figure}

In order to describe experimental observations, the details of theoretical model are presented as follows.
	As shown in the Fig. \ref{figm1}, the transmission of the light traverses three medium: the air, the substrate, and the metasurface.
	
	We first consider light transmission through the air/substrate  interface. Refraction angle satisfies ordinary Snell's law: $\frac{\sin(\theta'')}{\sin(\theta)}=\frac{n_0}{n_i}$ where $n_0$ and $n_i$ are the refractive indexes of air and substrate, respectively.
	The $x$ and $y$ components of the electric field in the substrate satisfy the Fresnel formula
	$$E_x=\frac{2\cos(\theta)^2}{\sqrt{2}[\cos(\theta)+n_2\cos(\theta'')]}, \quad E_y=\frac{i 2 \cos(\theta)}{\sqrt{2}[\cos(\theta'')+n_2\cos(\theta)]}.$$
	
	After passing through the substrate light interacts with the metasurface where the refraction is described by the generalized Snell's law and the amplitude obeys Eq. (10) in the main text.  In order to model the structural birefringence of the rectangular nano pillars we define the x nd y components of the susceptibility tensor of GaN nanopillar according to already defined after Eq. (S8). This functional form of the susceptibility Cartesian components with two resonant  frequencies $\omega_0$ and $\omega_1$ \cite{Houdre}, which agrees with angle resolved photoluminescence experiments of the birefringent material. In addition, by expanding the angle-dependent susceptibility components in a Taylor series, one can obtain an angular dispersion similar to that observed in THz metasurfaces \cite{Qiu}, where the phenomenological amplitudes in the expansion represent inter-particle coupling strengths. The transmission through the metasurface is further described by our mesoscopic model (see Eq. (10)), which allows us to obtain the final anomalous refraction efficiency.

\section{Design of the polarization dependent beam splitter}

The interferometric measurement between normal and anomalous refraction imposes that the scattering properties of the interface should contain both zero and first diffraction order with similar amplitude and polarization. The latter can be addressed by inserting a quarter wave plate into the path of the one of the diffracted beams. The former condition requires tuning the antenna scattering parameter. The amplitude and phase responses of the nanopillars forming the PB metasurface are related to the length and width of the nanopillar with a constant height of 800 nm. To maximize the PB metasurface efficiency, the antenna should maximally convert the polarization from the opposite orthogonal circular polarization, thus introducing a phase shift between ordinary and extraordinary axis of a nanopillar \cite{34}. However, in contrast to the previous PB metasurface works seeking for the high performance devices \cite{Qiao_2016}, we are herein interested in quantifying the PB phase from the self-referenced interferometric measurements. Proper design is achieved by considering birefrigent nanopillar introducing $\pi/2$ or $3\pi/2$ phase retardation, as indicated in Fig. 2A. To account for both tapering and to compensate for anisotropic etching, we choose the nanopillar size corresponding to $3\pi/4$ phase shift. To quantify the phase-shift of the light transmitted through the GaN nanopillars, we perform the electromagnetic simulations of the light transmission through a subwavelength array of nanopillars arranged in a square lattice.
The metasurface consists of GaN nanopillar array on Sapphire (\ce{Al2O3}) substrate. The refractive index for GaN has the following Sellmeier like relation \cite{gan}:
$$
n_{0}(\lambda)=\sqrt{1+\frac{A_0\lambda^2}{\lambda^2-(\lambda_0^G)^2}
+\frac{B_0\lambda^2}{\lambda^2-(\lambda_0^H)^2}},$$
$$
n_{e}(\lambda)=\sqrt{1+\frac{A_e\lambda^2}{\lambda^2-(\lambda_e^G)^2}
+\frac{B_e\lambda^2}{\lambda^2-(\lambda_e^H)^2}}.$$
Here, $A_0=0.213$, $B_0=3.988$, $A_e=0.118$, $B_e=4.201$, $\lambda_0^G=\lambda_e^G=350nm$, $\lambda_0^H=153nm$, $\lambda_e^H=173.5nm$. For Sapphire (\ce{Al2O3}), refractive index relation is referred from \cite{Sapphire}. For $\lambda=632.8$ nm, the design wavelength, $n=1.766$.  GaN has very small birefringence, which we decided to neglect as a first approximation.
To avoid diffraction both in free space and in the substrate, we arranged the spacing between the elements with a subwavelength period of $320$ nm. The simulations are performed using the FDTD using plane wave sources at $632.8nm$ wavelength polarized along $x$ and $y$ axis, impinging at normal incidence satisfying the perfectly matched layer (PML) conditions in the direction of the light propagation subject to periodic boundary conditions along all the in-plane directions. The use of PML boundary conditions in the propagation direction results in an open space simulation while in-plane periodic boundary conditions mimic a subwavelength array of the identical nanostructures.

\begin{figure}[ht]
	\centering
	\includegraphics[scale=0.6]{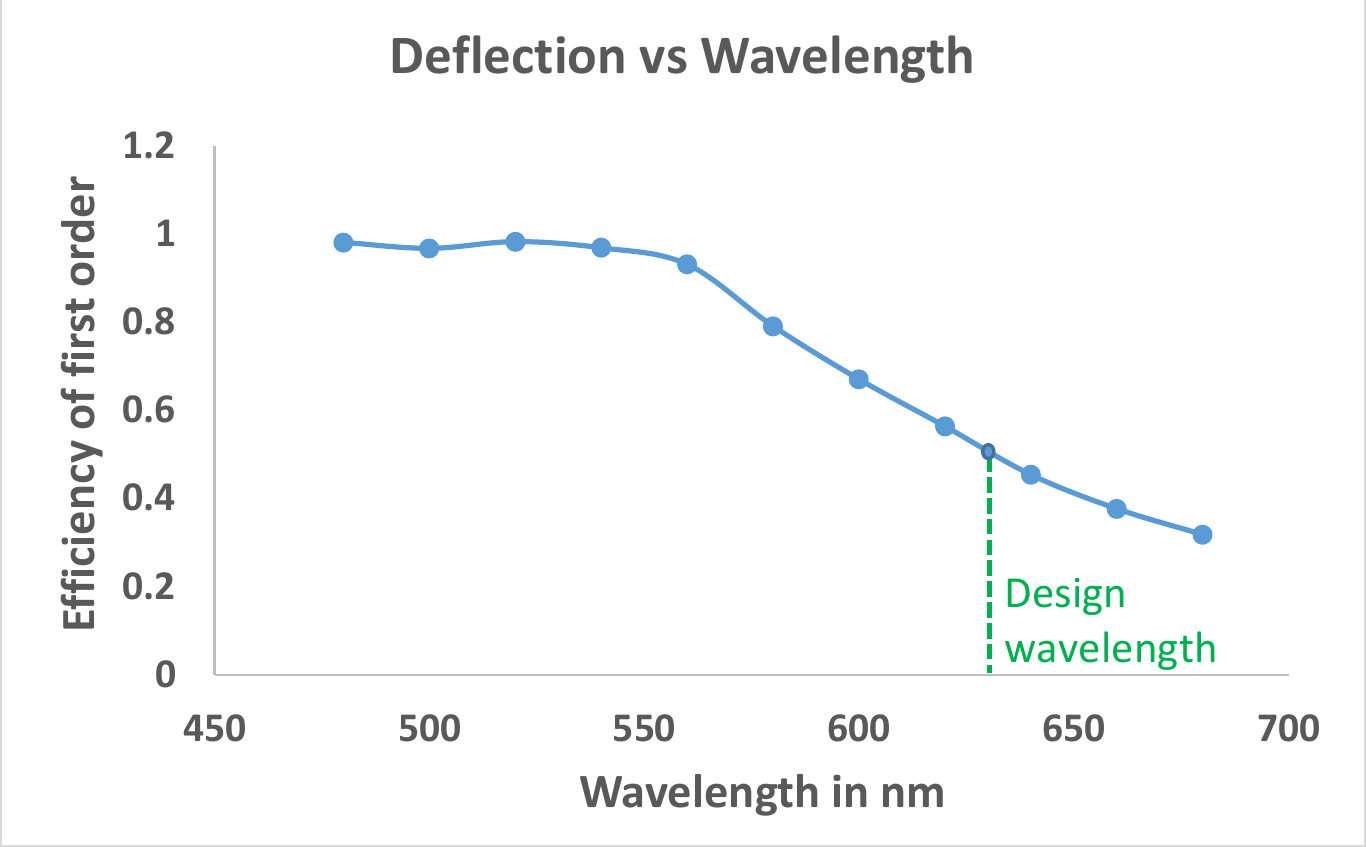}
	\caption{\footnotesize Polarization conversion efficiency of a single GaN nanopillar.}
	\label{figm2}
\end{figure}

In figure \ref{figm2}, the polarization-conversion efficiency of a single GaN nanopillar is obtained by FDTD simulation. A meta-atom is impinged with two plane waves sources (x and y-polarized) of varying wavelength from 480 nm to 680 nm with the interval of 20 nm. The other simulation conditions are kept same as in the design simulation. To perform interference measurements as described in the main text, the design of the element is chosen to  diffract  $50\%$ of the incident light on cross polarization.

\section{Interpretation of the interferometric experiments}

Note, that Eq. (12) shows that MZI detects the displacement phase. Technically the displacement phase is a propagation phase, also known as detour phase, since the measurement involves translational motion alone. However as has been already pointed out in the discussion following Eq. (7), the propagation phase contains two parts, one of which cancels out the PB phase yielding non vanishing first diffraction order. The detour phase is therefore a remaining part of the propagation phase which enters the Snell's law. Note, however, that due to the translational invariance and uniform distribution of PB phases in the unit cell, the magnitude of the displacement phase is equivalent to the PB phase. This equivalence is not accidental and follows directly from the metasurface design itself, and therefore can be controlled at will.  The MZI measurements thus provides although indirect yet unambiguous and conclusive measurement of the PB phase.

\begin{figure}[t]
	\centering
	\includegraphics[scale=0.25]{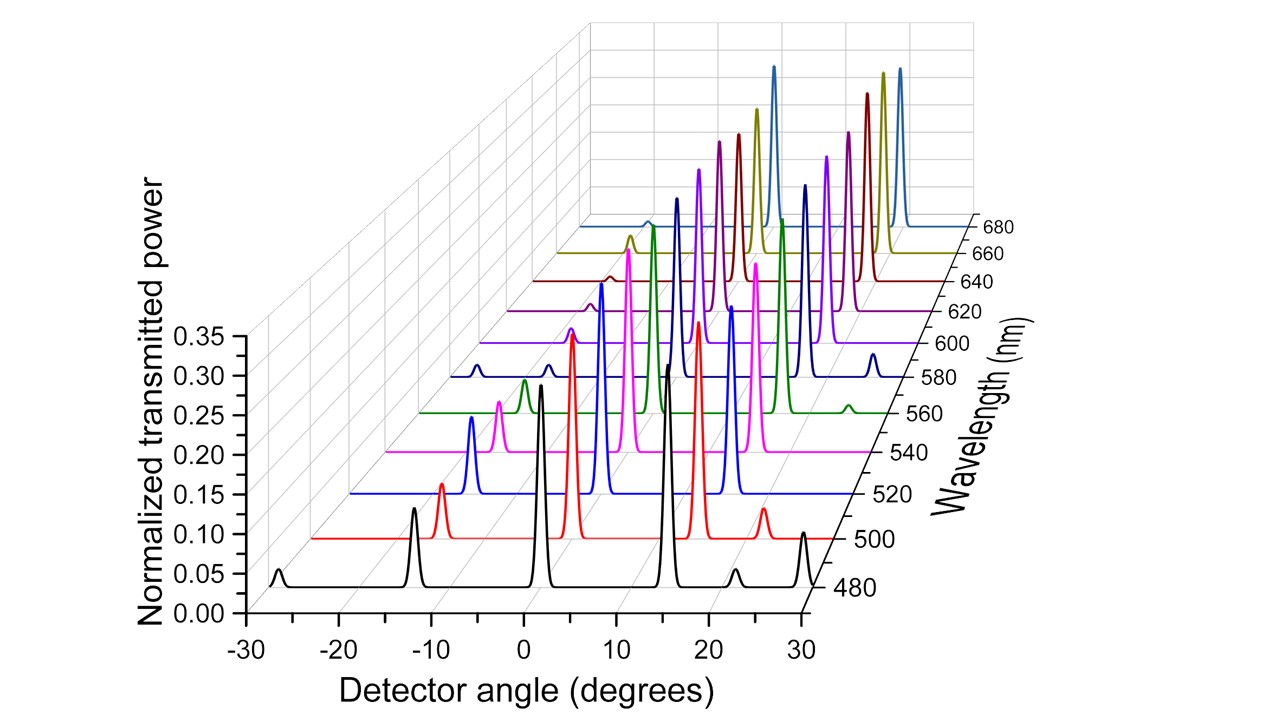}
	\includegraphics[scale=0.25]{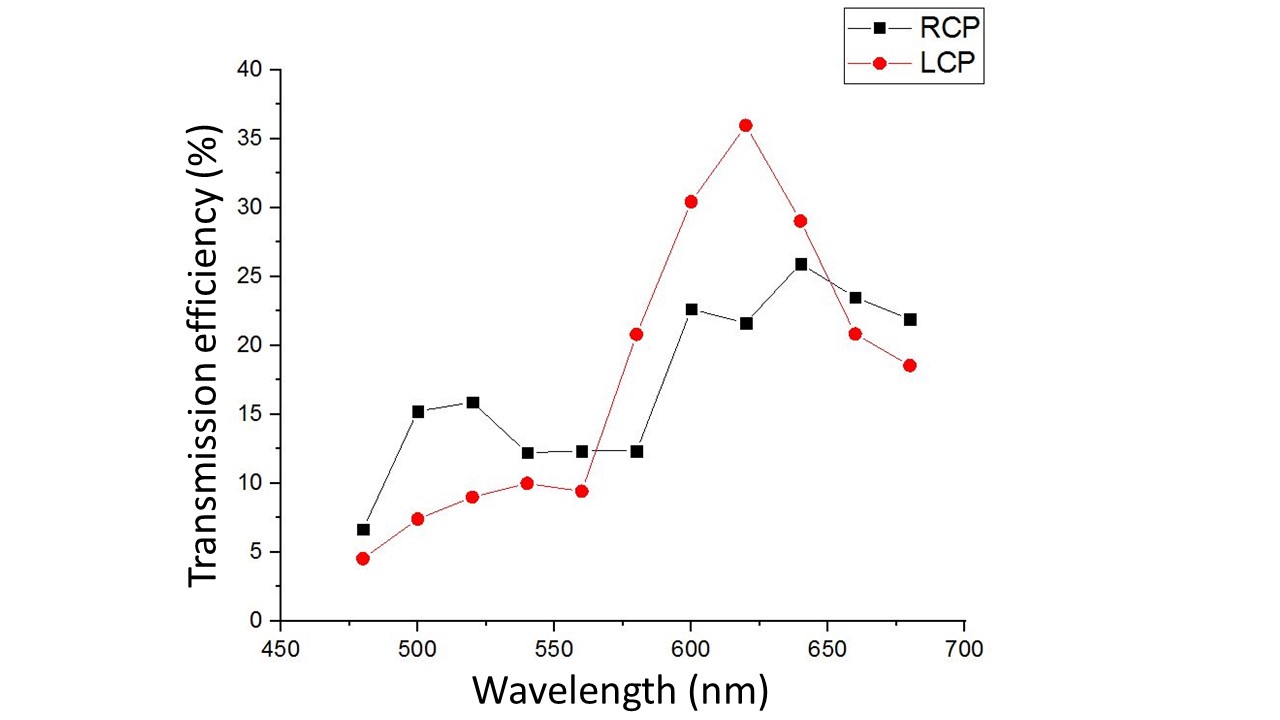}
	\caption{\footnotesize left: Measured angular deflection efficiency as a function of the incident wavelength for $\sigma_+$ incident polarization, to be compared with Fig 2D. Right: deflection efficiency as a function of the wavelength.}
	\label{fig2}
\end{figure}